\documentclass[twocolumn,times,tighten,trackchanges]{aastex63}

\usepackage{natbib}
\usepackage{xspace}
\usepackage{amsmath}
\usepackage{url}
\usepackage{appendix}

\newcommand{\lamlam}{$\lambda\lambda$}
\newcommand{\HII}{\mbox{H\,{\sc ii}}}

\newcommand{\CII}{\mbox{C\,{\sc ii}}}

\newcommand{\OIII}{\mbox{O\,{\sc iii}}}
\newcommand{\NII}{\mbox{N\,{\sc ii}}}
\newcommand{\NIII}{\mbox{N\,{\sc iii}}}
\newcommand{\Ha}{\mbox{H$\alpha$}}
\newcommand{\Hb}{\mbox{H$\beta$}}
\newcommand{\Lya}{\mbox{Ly$\alpha$}}

\newcommand{\SFR}{\ensuremath{\mathrm{SFR}}}

\newcommand{\kms}{$\mathrm{km\ s^{-1}}$}
\newcommand{\Mo}{$\mathrm{M_\sun}$}
\newcommand{\Moyr}{$\mathrm{M_\sun\ yr^{-1}}$}

\newcommand{\Lo}{$\mathrm{L_\sun}$}

\newcommand{\Zo}{$\mathrm{Z_\sun}$}

\newcommand{\um}{$\mathrm{\mu m}$}

\newcommand{\uJy}{$\mathrm{\mu Jy}$}
\newcommand{\mJykms}{$\mathrm{mJy\ km\ s^{-1}}$}
\newcommand{\Jykms}{$\mathrm{Jy\ km\ s^{-1}}$}
\newcommand{\uJybeam}{$\mathrm{\mu Jy\ beam^{-1}}$}

\newcommand{\mJybeamkms}{$\mathrm{mJy\ beam^{-1} km\ s^{-1}}$}
\newcommand{\uJybeamkms}{$\mathrm{\mu Jy\ beam^{-1} km\ s^{-1}}$}

\newcommand{\cmmm}{\ensuremath{\mathrm{cm^{-3}}}}

\newcommand{\NO}{\ensuremath{\text{N}/\text{O}}}
\newcommand{\logNO}{\ensuremath{\log_{10}(\text{N}/\text{O})}}
\newcommand{\OH}{\ensuremath{\text{O}/\text{H}}}
\newcommand{\logOH}{\ensuremath{\log_{10}(\text{O}/\text{H})}}
\newcommand{\LNii}{\ensuremath{L_\text{[NII]122}}}
\newcommand{\LOiii}{\ensuremath{L_\text{[OIII]88}}}
\newcommand{\LCii}{\ensuremath{L_\text{[CII]158}}}
\newcommand{\LHa}{\ensuremath{L_\mathrm{H\alpha}}}
\newcommand{\LNiiLOiii}{\ensuremath{L_\text{[NII]122}/L_\text{[OIII]88}}}
\newcommand{\CPDR}{\ensuremath{C_\text{PDR}}}
\newcommand{\SN}{\ensuremath{ \text{S}/\text{N}}}

\newcommand{\nn}{\mbox{--}}

\accepted{September 24, 2021}

\shorttitle{Big Three Dragons: [N {\sc ii}] 122 and dust-continuum observations}
\shortauthors{Sugahara et al.}

\begin{document}

\title{Big Three Dragons: A [N {\sc ii}] 122 \micron\ Constraint and New Dust-continuum Detection of A $z = 7.15$ Bright Lyman Break Galaxy with ALMA}

\author[0000-0001-6958-7856]{Yuma Sugahara}
\email{sugayu@aoni.waseda.jp}
\affil{National Astronomical Observatory of Japan, 2-21-1 Osawa, Mitaka, Tokyo 181-8588, Japan}
\affil{Waseda Research Institute for Science and Engineering, Faculty of Science and Engineering, Waseda University, 3-4-1, Okubo, Shinjuku, Tokyo 169-8555, Japan}
\author[0000-0002-7779-8677]{Akio K. Inoue}
\affil{Waseda Research Institute for Science and Engineering, Faculty of Science and Engineering, Waseda University, 3-4-1, Okubo, Shinjuku, Tokyo 169-8555, Japan}
\affil{Department of Physics, School of Advanced Science and Engineering, Faculty of Science and Engineering, Waseda University, 3-4-1, Okubo, Shinjuku, Tokyo 169-8555, Japan}
\author[0000-0002-0898-4038]{Takuya Hashimoto}
\affil{Tomonaga Center for the History of the Universe (TCHoU), Faculty of Pure and Applied Sciences, University of Tsukuba, Tsukuba, Ibaraki 305-8571, Japan}
\author[0000-0002-7738-5290]{Satoshi Yamanaka}
\affil{Waseda Research Institute for Science and Engineering, Faculty of Science and Engineering, Waseda University, 3-4-1, Okubo, Shinjuku, Tokyo 169-8555, Japan}
\affil{Research Center for Space and Cosmic Evolution, Ehime University, 2-5 Bunkyo-cho, Matsuyama, Ehime 790-8577, Japan}
\author[0000-0001-7201-5066]{Seiji Fujimoto}
\affil{Cosmic Dawn Center (DAWN), Jagtvej 128, DK2200 Copenhagen N, Denmark}
\affil{Niels Bohr Institute, University of Copenhagen, Lyngbyvej 2, DK2100 Copenhagen, Denmark}
\author[0000-0003-4807-8117]{Yoichi Tamura}
\affil{Division of Particle and Astrophysical Science, Graduate School of Science, Nagoya University, Nagoya 464-8602, Japan}
\author[0000-0003-3278-2484]{Hiroshi Matsuo}
\affil{National Astronomical Observatory of Japan, 2-21-1 Osawa, Mitaka, Tokyo 181-8588, Japan}
\affil{The Graduate University for Advanced Studies (SOKENDAI), 2-21-1, Osawa, Mitaka, Tokyo 181-8588, Japan}
\author[0000-0002-0808-4136]{Christian Binggeli}
\affil{Observational Astrophysics, Department of Physics and Astronomy, Uppsala University, Box 516, SE-751 20 Uppsala, Sweden}
\author[0000-0003-1096-2636]{Erik Zackrisson}
\affil{Observational Astrophysics, Department of Physics and Astronomy, Uppsala University, Box 516, SE-751 20 Uppsala, Sweden}

\begin{abstract}
We present new Atacama Large Millimeter/submillimeter Array  Band 7 observational results of a Lyman break galaxy at \(z=7.15\), B14-65666 (``Big Three Dragons''), which is an object detected in [\OIII] 88 \um, [\CII] 158 \um, and dust-continuum emission during the epoch of reionization.
Our targets are the [\NII] 122 \um\ fine-structure emission line and underlying 120 \um\ dust continuum.
The dust continuum is detected with a \(\sim\)19\(\sigma\) significance.
From far-infrared spectral energy distribution sampled at 90, 120, and 160 \um, we obtain a best-fit dust temperature of \(40\) K (\(79\) K) and an infrared luminosity of \(\log_{10}(L_{\rm IR}/{\rm L}_\odot)=11.6\) (\(12.1\)) at the emissivity index \(\beta = 2.0\) (1.0).
The [\NII] 122 \um\ line is not detected.
The 3\(\sigma\) upper limit of the [\NII] luminosity is \(8.1 \times 10^7\) \Lo.
From the [\NII], [\OIII], and [\CII] line luminosities, we use the Cloudy photoionization code to estimate nebular parameters as functions of metallicity.
If the metallicity of the galaxy is high (\(Z > 0.4\) \Zo), the ionization parameter and hydrogen density are \(\log_{10} U \simeq -2.7\pm0.1\) and \(n_\text{H} \simeq 50\nn250\) \cmmm, respectively, which are comparable to those measured in low-redshift galaxies.
The nitrogen-to-oxygen abundance ratio, \NO, is constrained to be sub-solar.
At \(Z < 0.4\) \Zo, the allowed \(U\) drastically increases as the assumed metallicity decreases.
For high ionization parameters, the \NO\ constraint becomes weak.
Finally, our Cloudy models predict the location of B14-65666 on the BPT diagram, thereby allowing a comparison with low-redshift galaxies.
\end{abstract}


\section{INTRODUCTION}
\label{sec:intro}
The Atacama Large Millimeter/submillimeter Array (ALMA) has contributed to several pioneering works on far-infrared (FIR) fine-structure lines in star-forming galaxies at redshift \(z > 7\), providing new insights into galaxy formation and evolution at the earliest epochs.
The [\CII] 158 \um\ emission line, one of the brightest emission lines in the FIR band, is commonly used to trace high-redshift galaxy properties \citep[e.g.,][]{Capak:2015,Carniani.S:2018a, Le-Fevre.O:2020a,Bakx.T:2020a}, even at \( z>7 \) \citep{Maiolino.R:2015a,Pentericci.L:2016a,Hashimoto.T:2019a,Carniani.S:2020a}.
While the [\CII] line is a dominant coolant in neutral gas \citep{Tielens.A:1985a, Abel.N:2005a} and relevant to star-forming activities \citep[e.g.,][]{Boselli.A:2002a, De-Looze.I:2014a, Schaerer.D:2020a, Fujimoto.S:2021a}, the low ionization potential of C$^+$ (11.3 eV) permits [\CII] emission in various phases: \HII\ regions, photodissociated regions (PDRs), cold neutral and molecular medium, and shocks caused by galaxy interactions \citep[e.g.,][]{Russell.R:1980a, Tielens.A:1985a, Appleton.P:2013a}.
The [\OIII] 88 \um\ is another tracer for high-redshift star formation \citep{Inoue.A:2014a} and has also been detected at \( z\gtrsim7 \) \citep{Inoue.A:2016a,Carniani.S:2017b,Laporte.N:2017a,Marrone.D:2018a,Hashimoto.T:2018a,Hashimoto.T:2019a,Tamura.Y:2019a}.
In contrast to the [\CII] line, the [\OIII] line arises only from \HII\ regions due to high ionization potential of O$^{2+}$ (35.1eV).
The combination of these two FIR lines provides information on the physical conditions of the interstellar medium (ISM) in individual high-redshift galaxies, which are difficult to probe with weak nebular emission lines in the rest-frame ultraviolet (UV) band.
\citet{Inoue.A:2016a} found a deficit in [\CII]-to-[\OIII] luminosity ratios compared with local galaxies, possibly suggesting a highly ionized state in high-redshift galaxies.
This is supported by recent observational studies (\citealp{Hashimoto.T:2019a,Harikane.Y:2020b}; but see \citealp{Carniani.S:2020a})

In addition to fine-structure lines, high sensitivity observations with ALMA enable the detection of FIR continuum emission at \( z > 7 \) \citep{Watson.D:2015a, Laporte.N:2017a, Hashimoto.T:2019a, Tamura.Y:2019a}.
Infrared luminosity is dominated by thermal dust emission, reflecting UV energy absorbed by dust.
Therefore, FIR spectral energy distribution (SED) is a clue for constraining dust properties, including dust temperature and dust mass.
\citet{Bakx.T:2020a} reported the non-detection of dust continuum at \( 160 \) \um\ in a Lyman break galaxy (LBG) at \( z=8. 3113\), MACS0416\_Y1, despite detecting it at \( 90 \) \um.
These results suggest an unusually high dust temperature \( T_\text{dust}>80 \) K or a high emissivity index \( \beta > 2 \).
While FIR SED is usually fitted by a modified blackbody function with an assumed dust temperature, \citet{Inoue.A:2020a} proposed a new algorithm for determining dust temperature based on radiative equilibrium on dust grains.

This paper presents new observations of an LBG, B14-65666, which is the first example of a galaxy detected in [\OIII] 88 \um, [\CII] 158 \um, and underlying dust continuum at the epoch of reionization \citep[so-called ``Big Three Dragons,''][]{Hashimoto.T:2019a} in order to take a step further in understanding galaxy properties at high-redshift.
The new observations are used to detect another FIR emission line [\NII] 122 \um\ (2459.380 GHz\footnote{This is taken from the Spectral Line Atlas of Interstellar Molecules (SLAIM) \citep[Available at \url{http://www.splatalogue.net}; F. J. Lovas, private communication,][]{Remijan.A:2007a}.}) and underlying dust continuum at 120 \um.
An additional dust continuum measurement is useful for constraining dust properties.
Observations of [\NII] lines are limited at high-redshift, but the number of detections has been increasing.
[NII] 122 \um\ lines are detected in quasars at \(z=2.56\) \citep{Ferkinhoff.C:2011a}, \( 6.003 \) \citep{Li.J:2020b}, and \( 7.54\) \citep{Novak.M:2019a}, a lensed dusty star-forming galaxy (DSFG) at \( z = 2.3 \) \citep{George.R:2014a}, \( 2.81 \) \citep{Ferkinhoff.C:2011a, Ferkinhoff.C:2015a} and \( 4.22 \) \citep{De-Breuck.C:2019a}, and a submillimeter galaxy (SMG) and quasar system at \( z=4.69 \) \citep{Lee.M:2019a}, whereas \citet{Harikane.Y:2020b} reported three LBGs at \( z\sim6 \) that are not detected in [\NII] 122 \um\ lines despite detecting them in [\CII] and [\OIII] lines.
In another excitation state, [\NII] 205 \um\ lines are detected in galaxies at \( z=5\nn6 \) \citep{Pavesi.R:2016a}, SMGs at \( 3<z<6 \) \citep{Cunningham.D:2020a}, and \( z>4 \) objects, including quasars and DSFGs \citep[e.g.,][]{Decarli.R:2014a, De-Breuck.C:2019a, Novak.M:2019a, Cheng.C:2020a}.

It is important to study galaxy ISM with emission-line diagnostics thorough a wide range of redshifts at \( z=0\nn7 \).
At \( z=0\nn2 \), physical properties in star-forming galaxies are probed in detail using rest-frame optical emission lines.
The optical wavelength range includes emission lines from different elements (e.g., H, O, N, S), excited states (e.g., [\NII]\lamlam6549,6585), and ionization states (e.g., O$^{+}$ and O$^{2+}$); thus, temperatures, densities, elemental abundances, and ionizing sources in \HII\ regions can be inferred by comparing emission-line fluxes.
The nitrogen-to-oxygen (\NO) abundance ratio provides a clue to the chemical evolution in galaxies \citep[e.g.,][]{Vincenzo.F:2016a} because more nitrogen is produced through the carbon-nitrogen-oxygen (CNO) cycle as a secondary nucleosynthesis product in stars with higher metallicity.
Among dominant ionizing sources of strong line emitters, star-formation and active galactic nuclei (AGNs) can be distinguished on the [\OIII]\(/\)\Hb--[\NII]\(/\)\Ha\ plane, the so-called Baldwin-Phillips-Terlevich (BPT) diagram \citep{Baldwin.J:1981a,Veilleux.S:1987a}.
The distribution on the BPT diagram is theoretically interpreted as a combination of the nebular physical parameters and hardness of ionizing radiation \citep{Kewley.L:2013b}.

Galaxy properties at \( z=2 \) have been explored using near-infrared (NIR) instruments to detect redshifted optical lines.
On the BPT diagram, the sequence of star-forming galaxies at \( z\sim2 \) has an offset from the local sequence \citep[e.g.,][]{Shapley.A:2005b, Erb:2006a}, which is statistically confirmed using the KBSS-MOSFIRE sample \citep{Steidel:2014, Steidel:2016, Strom.A:2017b}, the MOSDEF sample \citep{Shapley.A:2015a, Sanders.R:2016a, Shivaei.I:2018a}, and Subaru/FMOS observations \citep{Yabe.K:2014a, Hayashi.M:2015a, Kashino.D:2017a}.
This offset on the BPT diagram originates from the redshift evolution of ionization states in star-forming galaxies.
\citet{Strom.A:2018a} derived the \NO\ abundance ratios of KBSS-MOSFIRE galaxies to find that the \NO\ ratio at \( z\sim2 \) is comparable with local abundance ratios \citep{Pilyugin.L:2012a} at fixed metallicities even though an increase in the \NO\ ratio is suggested from local to \( z\sim2 \) in some studies \citep[e.g.,][]{Masters.D:2014a, Sanders.R:2016a, Kojima.T:2017a}.

The optical and NIR observations are powerful tools for galaxy properties, but a redshift range is limited from \( z\sim0 \) to \( 2 \).
We have to wait for MIR observations with the James Webb Space Telescope (JWST) to extend the studies for higher redshifts.
On the other hand, few galaxies are detected in FIR fine-structure lines including [\OIII] 88 \um\ at \( 0 < z < 3 \) \citep[e.g.,][]{Ferkinhoff.C:2010a, Zhang.Z:2018a}, because FIR observations of fine-structure lines are mainly available for galaxies at \( z\gtrsim 2\nn3 \) using ALMA or for local galaxies using telescopes such as Herschel, ISO, AKARI, and SOFIA.
This redshift gap in FIR observations makes it difficult to discuss a continuous galaxy evolution scenario.
We propose comparisons of the physical ISM properties at various redshifts estimated from either FIR or optical emission lines to overcome this difficulty, with the aid of a photoionization model.

This paper consists of the following sections.
Section \ref{sec:sample} describes the B14-65666 observations and data reduction process.
Section \ref{sec:results} presents the dust continuum and [\NII] 122 \um\ emission line measurements.
Section \ref{sec:FIR-SED-fitting} explains FIR SED fittings and their results.
We evaluate nebular parameters from [\OIII] 88 \um\ and [\CII] 158 \um\ in Section \ref{sec:discussion}.
We discuss the \NO\ abundance ratio and BPT diagram at \( z\sim7 \) using these nebular parameters.
Section \ref{sec:conclusion} summarizes our findings.

\begin{figure*}[t]
    \epsscale{1.1}
    \plotone{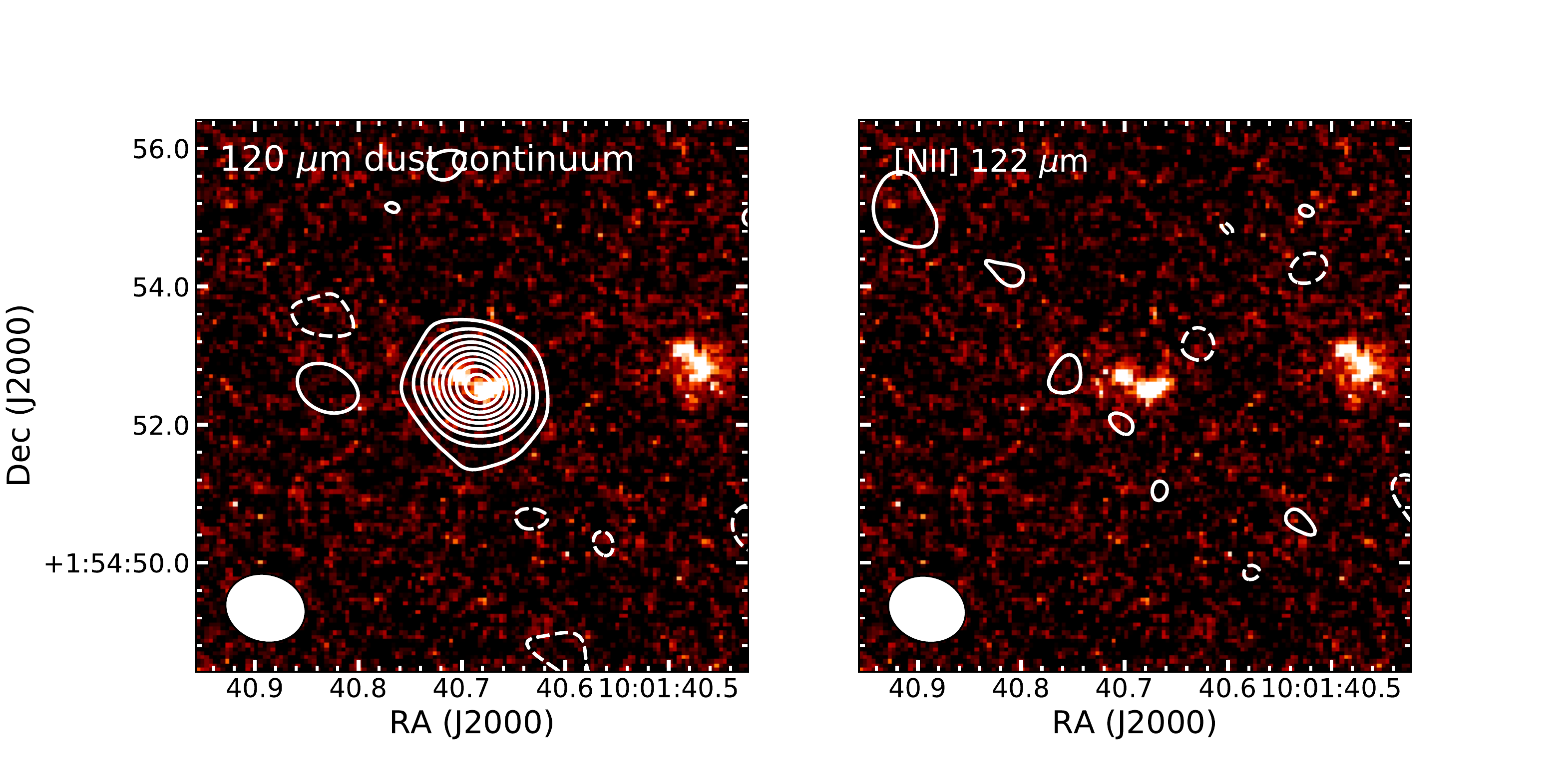}
    \caption{
      Left: ALMA 120 \um\ dust contours overlaid on the HST/WFC3 \textit{F140W}-band pseudo-color image.
      The contours are drawn with \( 2\sigma \) intervals from the \( -2\sigma \) to \( 18\sigma \) level, where \( \sigma = 9.8\) \uJybeam.
      The white ellipse at the lower left shows the size of the synthesized beam.
      Right: ALMA [\NII] 122 \um\ contours overlaid on the HST/\textit{F140} image.
      The integrated velocity range for the contours is from \( -201 \) to \( 203 \) \kms\  with respect to the systemic redshift measured using [\OIII] and [\CII].
      The contours are drawn with \( -2\sigma \) and \( 2\sigma \) levels, where \( \sigma = 17.1 \) \uJybeamkms.
      As seen, the [\NII] 122 \um\ emission line is not detected.
    }
  \label{fig:10}
\end{figure*}

\section{Observation and Data Reduction}
\label{sec:sample}
Our target object, B14-65666, is located at RA 10$^\text{h}$01$^\text{m}$40\fs69, Dec. \(+\)01\degr 54\arcmin52\farcs42 (J2000).
It was found by \citet{Bowler.R:2014a} and spectroscopically detected in \Lya\ by \citet{Furusawa.H:2016a}.
\citet{Bowler.R:2018a} performed ALMA follow-up observations with Band 6 in Cycle 3 and reported the detection of 160 \um\ dust continuum.
\citet{Hashimoto.T:2019a} detected [\OIII] 88 \um, [\CII] 158 \um, and underlying dust continua in Cycles 4 and 5.

We target the [\NII] 122 \um\ emission line, which is free from strong atmospheric absorption, to obtain a signature of nitrogen.
The frequency (wavelength) of the line at the rest frame is 2459.380 GHz (121.898 \um).
An advantage of this line is it has a similar critical density to the [\OIII] 88 \um\ (\( 238 \) and \( 500 \) \cmmm\ at \(10^4\) K, respectively\footnote{In this paper, the critical density for a given excited state is defined as the density at which the sum of collisional excitation and de-excitation rates balances the spontaneous emission rate. Critical densities are computed using PyNeb \citep{Luridiana.V:2015a}.}), resulting in a weak dependence of the [\NII] 122 \um\( / \)[\OIII] 88 \um\ line ratio on electron density.
In contrast, the [\NII] 205 \um\ line has a lower critical density (\( 38 \) \cmmm) by an order of magnitude than the [\OIII] line, leading to a quick drop of the [\NII]-to-[\OIII] line ratio at high density.
The [\NIII] 57 \um\ line is another candidate observable with Band 9.
Since \(\text{N}^{2+}\) has a similar ionization potential to \( \text{O}^{2+} \), [\NIII]-to-[\OIII] line ratios do not depend on the ionization parameter as strongly as [\NII]-to-[\OIII] ratios, but it takes a longer integration time to detect the [\NIII] 57 \um\ line with the same significance level as the [\NII] 122 \um\ line.

We have observed B14-65666 in 2019 November with ALMA Band 7 during Cycle 7 (ID: 2019.1.01491.S, PI: A. K. Inoue).
The 12-m array was configured in the C43-2 configuration.
The correlator operated in a time division mode with 2.000 GHz bandwidths and 31.2 MHz spectral resolution.
One of the four spectral windows targets the [NII] 122 \um\ emission line at an expected frequency of 301.6885 GHz, and the others target dust continuum emission at 299.893, 289.500, and 287.700 GHz.
The total on-source exposure time was 143 minutes.
The bandpass/flux calibrators are quasars J1058+0133 and J0725-0054.
The phase calibrator is J1010-0200.

Data reduction and calibration are performed using a standard pipeline on Common Astronomy Software Applications \citep[CASA;][]{McMullin.J:2007a} version 5.6.1-8.
A dust continuum image is created using a CASA task \texttt{tclean} with natural weighting.
A line cube is created using \texttt{tclean} with a \( \simeq\!15.6 \) \kms\ spectral resolution after the dust continuum is subtracted with a CASA task \texttt{uvcontsub}.
In both the dust continuum image and line momentum 0 map (Section \ref{sec:nii-122-emission}), the beam size is approximately \( 1.15\arcsec \times 0.97\arcsec \), and the beam position angle is approximately \(  73^{\circ} \).

\section{Measurements}
\label{sec:results}

\subsection{120 \um\ Dust Continuum Emission}
\label{sec:dust-cont-emiss}
The left panel of Figure \ref{fig:10} illustrates the 120 \um\ dust continuum image with white contours overlaid on the \textit{F140W}-band image taken using the Wide Field Camera 3 (WFC3) on board the Hubble Space Telescope (HST).
The root mean square (rms) of the image is \( \sigma = 9.8\) \uJybeam.
The dust continuum is significantly detected with a peak signal-to-noise ratio (\SN) of \(18.9\).
This detection makes B14-65666 the second object in which dust continua are detected in more than two bands at \( z>7 \) after A1689zD1 \citep{Watson.D:2015a, Knudsen.K:2017a, Inoue.A:2020a}.
\citet{Bowler.R:2018a} suggested a physical offset of \( 3 \) kpc between the dust continuum and UV emission, but our measurement of 120 \um\ dust emission shows no spatial offset from UV emission on the WFC3/\textit{F140W} image, which has consistent astrometry with the images used by \citet{Bowler.R:2018a}.
Although 120 \um\ dust emission is not spatially resolved with this beam size, our result that no spatial offset was observed is consistent with the results on 90 and 160 \um\ dust continua reported by \citet{Hashimoto.T:2019a}.

We measure the spatially integrated flux density of the dust continuum using a 2D Gaussian profile with a CASA task \texttt{imfit}.
The measured flux density is \( 218 \pm 19 \) \uJy.
This flux density at 120 \um\ is less than the continuum flux density at 90 \um\ (\( 470\pm128 \) \uJy) but larger than at 160 \um\ (\( 130\pm25 \) \uJy), measured in \citet{Hashimoto.T:2019a}.
The 120 \um\ flux density has a higher \SN\ than the 90 and 160 \um\ flux density by a factor of 3--4.
These measurements are listed in Table \ref{tab:1}.
We examine the FIR SED with modified blackbody and radiative equilibrium models to estimate the IR luminosity based on \citet{Inoue.A:2020a} in Section \ref{sec:FIR-SED-fitting}.

\begin{figure}[t]
    \epsscale{1.0}
    \plotone{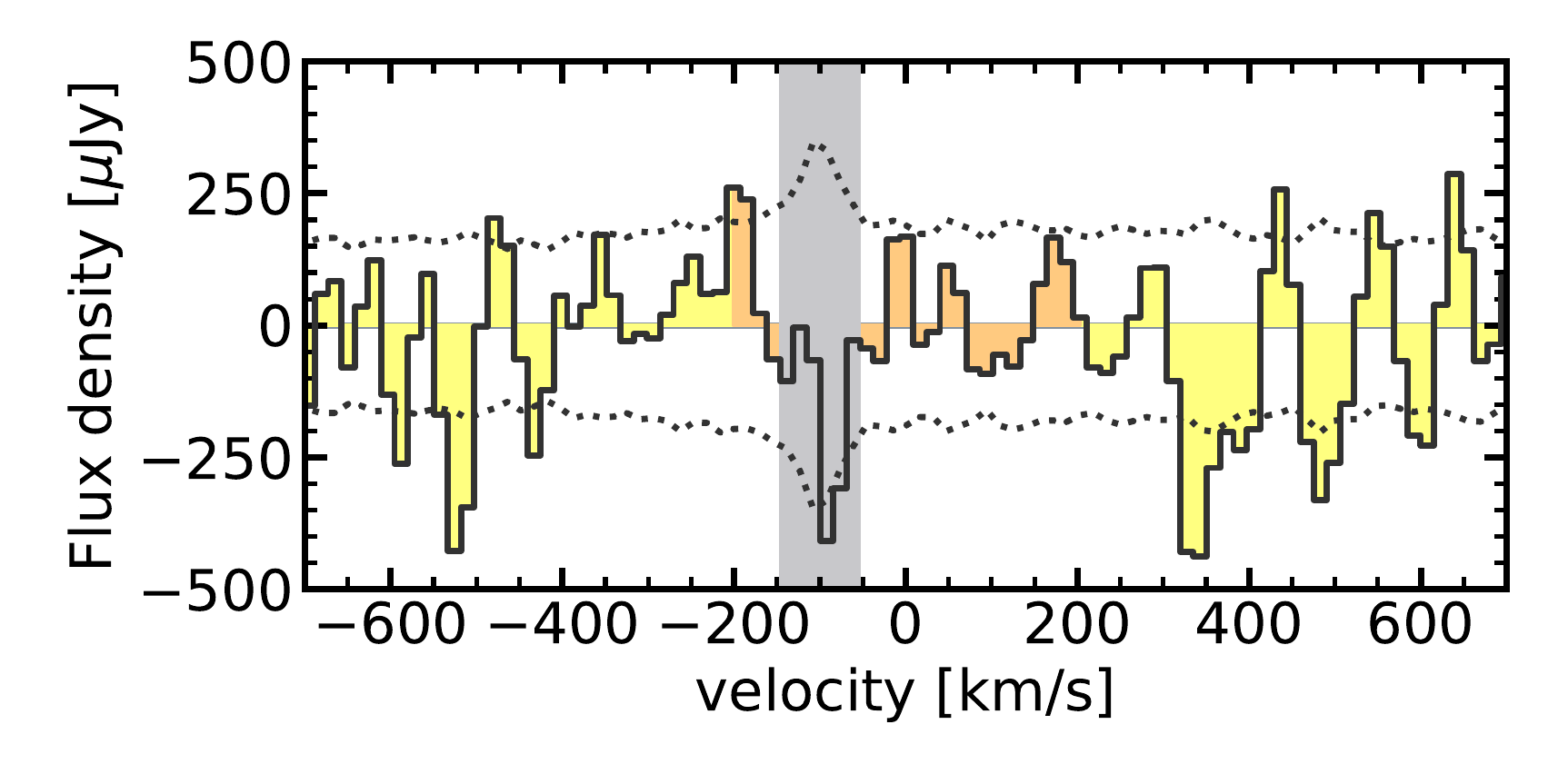}
    \caption{
      The spectrum around the [\NII] 122 \um\ line, where the velocity (optical definition) is relative to the systemic redshift of z=7.1521 determined by Hashimoto et al. (2019).
      The aperture radius for the spectrum is 1\arcsec.
      The dotted lines represent the noise spectrum.
      The range of velocities used for the [\NII] 122 \um\ line flux is highlighted in orange.
      The gray region is not used to measure the [\NII] line flux due to large spectral noise caused by weak atmospheric absorption.
      As seen, the [\NII] 122 \um\ emission line is not detected.
    }
  \label{fig:15}
\end{figure}

\begin{deluxetable}{lcc}
    \tablecaption{A summary of measurements of B14-65666.\label{tab:1}}
    \tablewidth{0pt}
    \tablecolumns{3}
    \tablehead{
    \colhead{Parameters} & \colhead{measurements} & \colhead{references}
    }
    \startdata
    {[\NII]} flux [\Jykms] & \( < 0.0514 \) & This study \\
    {[\OIII]} flux [\Jykms] & \( 1.5\pm0.18\) & H19 \\
    {[\CII]} flux [\Jykms] & \( 0.87\pm0.11 \) & H19 \vspace{3pt}\\
    {[\NII]} luminosity [\Lo] & \( <8.1\times10^7\) & This study \\
    {[\OIII]} luminosity [\Lo] & \( (3.4\pm0.4)\times10^9\) & H19 \\
    {[\CII]} luminosity [\Lo] & \( (1.1\pm0.1)\times10^9\) & H19 \vspace{3pt}\\
    90 \um\ flux density [\uJy] & \( 470\pm128 \) & H19 \\
    120 \um\ flux density [\uJy] & \( 218\pm19 \) & This study \\
    160 \um\ flux density [\uJy] & \( 130\pm25 \) & H19 \\
    \enddata
    \tablecomments{
      The measurements taken from H19 represent total values, meaning the whole emission from two clumps.
    }
    \tablerefs{
      H19: \citet{Hashimoto.T:2019a}
    }
\end{deluxetable}

\subsection{Upper Limit of [\NII] 122 \um\ Emission Line}
\label{sec:nii-122-emission}

We expect that the redshift of [\NII] emission lines will be the same as the redshift of [\OIII] emission lines because both of them, which have higher ionization potential than hydrogen, will arise from \HII\ regions.
We assume the observed-frame [\NII] 122 \um\ frequency as 301.7 GHz (993.7 \um) using a systemic redshift of \( z = 7.1521 \), which was determined from [\OIII] 88 \um\ and [\CII] 158 \um\ emission lines by \citet{Hashimoto.T:2019a}.

We find no emission line features around the observed-frame [\NII] frequency.
Assuming that the [\NII] line width is \( \simeq 400 \) \kms, which is close to the full-width half-maximum of [\CII] and [\OIII] lines \citep{Hashimoto.T:2019a}, we create a [\NII] flux (moment 0) map integrated from \(-201\) to \(203\) \kms\  around the observed-frame [\NII] frequency with a CASA task \texttt{immoments}, excluding channels at \( -139 \) to \( -61 \) \kms\ that are noisy through weak atmospheric absorption.
In the right panel of Figure \ref{fig:10}, the contours illustrate the [\NII] flux map.
Figure \ref{fig:15} shows a spatially-integrated spectrum in a 1\arcsec-radius aperture around the galaxy.
The noise spectrum shown by the dotted lines is measured by placing 500 random apertures of the same radii.
There is no significant detection with \SN\ \( > 3 \) in the image and spectrum.

An upper limit of the [\NII] line emission is measured from the uncertainty of the flux map.
The rms of the [\NII] flux map is \( 17.1 \) \mJybeamkms.
Given the source size of the HST image, the galaxy will not be resolved spatially.
Therefore, the \(3\sigma\) upper limit of the [\NII] flux is computed from the rms in the flux map to be \( 51.4 \) \mJykms, by adopting the spatial size of a single beam.
The line flux limit is converted to the line luminosity limit using the luminosity distance and observed frequency \citep{Carilli:2013}.
The [\NII] line luminosity is constrained to be \( 8.1 \times 10^7\) \Lo\ as the \( 3\sigma \) upper limit.
The [\NII] line flux and luminosity are listed in Table \ref{tab:1}, as well as the [\OIII] and [\CII] lines taken from \citet{Hashimoto.T:2019a}.

The constraint of [\NII] line luminosity is lower than the measurements for star-forming/starburst galaxies in literature.
\citet{Lee.M:2019a} detected an [\NII] 122 \um\ line in BRI 1202-0725 SMG at \( z= 4.96 \) and obtained the line luminosity of \( 2.71\pm0.65\times10^9\) \Lo.
The [\NII] line luminosity of SPT 0418-47 at \( z=4.2 \) measured by \citet{De-Breuck.C:2019a} is \( 1.6\pm0.5\times10^8 \) \Lo, which is corrected with the gravitational magnification factor of \( \mu=32.7 \) \citep{Spilker.J:2016a}.
SMMJ02399-0136 and the Cosmic Eyelash, lensed DSFGs at \( z = 2\nn3\), exhibit the lens-corrected [\NII] line luminosities of \( 2.0\pm0.3\times10^{10} \) \Lo\ \citep{Ferkinhoff.C:2015a} and \( 5.2\pm0.5\times10^{8} \) \Lo \citep{Zhang.Z:2018a}, respectively.
\citet{Harikane.Y:2020b} observed three LBGs at \( z\sim6 \) with ALMA but did not detect the [\NII] line for any of the three objects, while they did detect [\OIII] 88 \um\ and [\CII] 158 \um\ lines.
The [\NII] luminosities of these objects are \( < 6.2 \times 10^8 \), \( < 8.3\times10^8\), and \( < 1.2\times10^9\) \Lo\ for the upper limits, which are integrated in \( 600 \) \kms\ velocity width and a 2\arcsec-radius aperture.
Our upper limit of [\NII] luminosity is several times lower than these constraints in literature despite the higher redshift of B14-65666.

\section{FIR SED Fitting}
\label{sec:FIR-SED-fitting}

\subsection{Modified Blackbody Fitting}
\label{sec:mod-blackbody}

We perform FIR SED fitting to estimate the IR luminosity and dust mass by combining the new 120 \um\ dust continuum emission with previous measurements of 90 and 160 \um\ dust continua underlying [\OIII] 88 \um\ and [\CII] 158 \um.
In this paper, we only discuss the total FIR SED, assuming a constant dust temperature in the entire system of B14-65666, while the galaxy is composed of two components and suggested to be a merging system \citep{Hashimoto.T:2019a}.
First, we fit a standard modified blackbody function to the observed FIR flux densities: 
\begin{equation}
 F_\nu^{\rm obs}=\frac{1+z}{d_{\rm L}^2} M_{\rm d} \kappa_\nu \left\{B_\nu(T_{\rm d})-B_\nu(T_{\rm CMB})\right\}\,, 
 \label{eq:fnu}
\end{equation}
where $z$ is the source redshift, and $d_{\rm L}$ is the luminosity distance.
$M_{\rm d}$ is the dust mass, which is the normalization of the equation.
$\kappa_\nu$ is the dust emissivity at the frequency $\nu$.
$B_\nu(T)$ is the blackbody function, and $T_{\rm d}$ is the dust temperature of the source.
$T_{\rm CMB}$ is the cosmic microwave background (CMB) temperature at the source redshift, and $B_\nu(T_{\rm CMB})$ denotes CMB intensity.
The negative $B_\nu(T_{\rm CMB})$ term accounts for a correction of CMB effect in interferometric observations \citep{da-Cunha:2013}.

\begin{figure}[t]
    \epsscale{1.2}
    \plotone{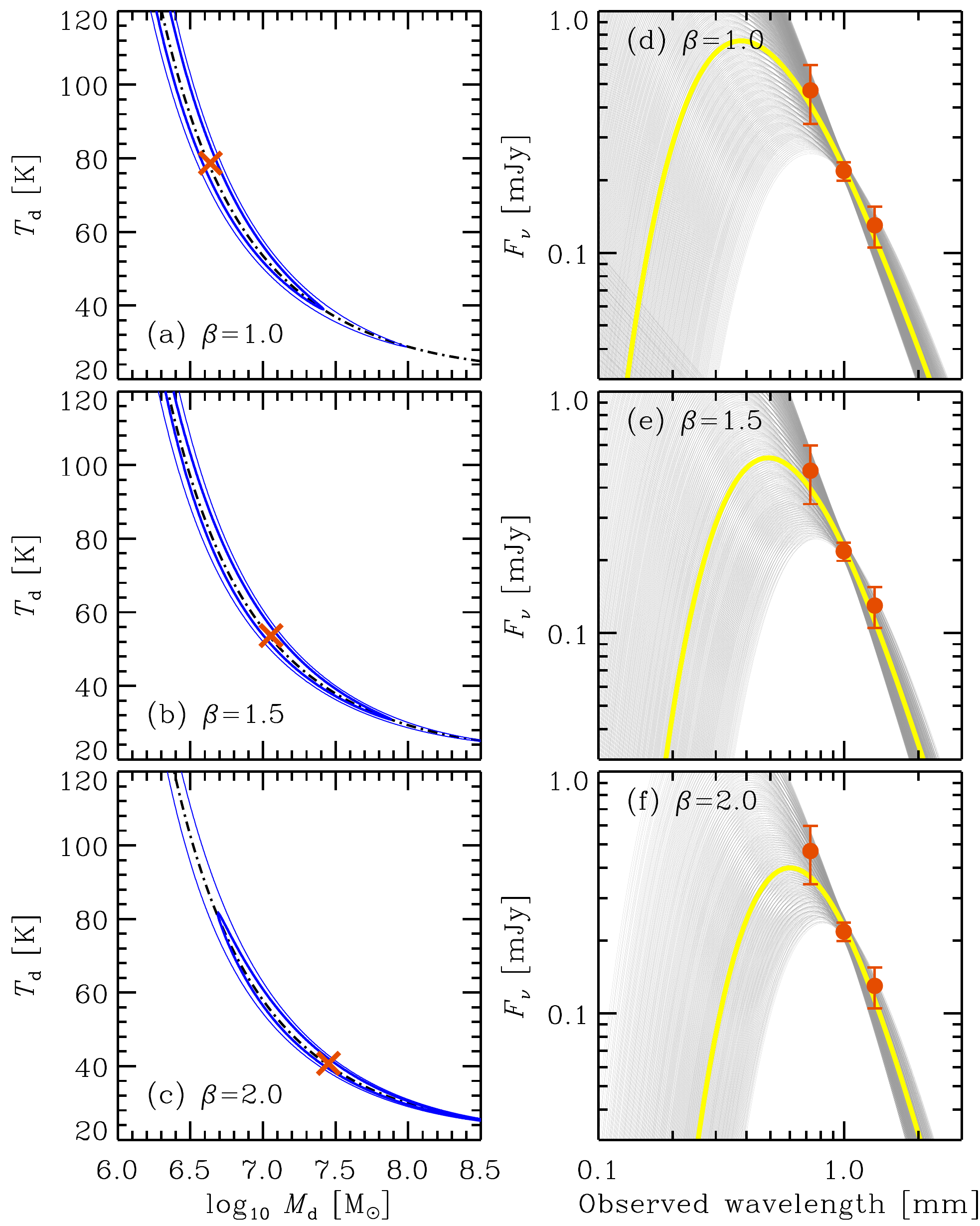}
    \caption{
        Modified blackbody fitting results. (a)--(c) The best-fit solutions (crosses) and central 68/95 percent (thick/thin solid lines) areas in the dust temperature and mass plane for the emissivity indices $\beta=1.0$, $1.5$, and $2.0$. The dot-dashed lines show the sequences of dust temperature and mass that provide the observed Band~7 flux density. (d)--(f) FIR SEDs for best-fit solutions (thick yellow solid lines) and those in the 68 percent areas of the dust temperature--mass plane for $\beta=1.0$, $1.5$, and $2.0$. The data points with error bars are the ALMA measurements.
    }
    \label{fig:mBBfit}
\end{figure}

Although assumptions on the emissivity value systematically affect estimates of dust mass \citep[e.g.,][]{Fanciullo.L:2020a}, there are large variations among empirical estimates, theoretical models, and laboratory measurements of emissivity, as briefly reviewed in \citet{Inoue.A:2020a}.
Following \citet{Inoue.A:2020a}, we assume a typical value of $\kappa_\nu=30~{\rm cm^2~g^{-1}}(100~{\micron}/\lambda)^\beta$ with the wavelength $\lambda=c/\nu$ and the light speed $c$ and the emissivity index $\beta=1.0$, $1.5$, or $2.0$.
The pivot value of \( 30\ \mathrm{cm^2~g^{-1}} \) at 100 \um\ is very similar to those of Astronomical silicate \citep{Draine.B:1984a, Weingartner.J:2001a} and the THMIS model \citep{Jones.A:2017a}.
Figure~\ref{fig:mBBfit} shows the results obtained from the least-$\chi^2$ fitting with two free parameters of $M_{\rm d}$ and $T_{\rm d}$.
There is a degeneracy between $M_{\rm d}$ and $T_{\rm d}$ because we do not constrain the peak of FIR SED yet.
A data point at a shorter wavelength than 90 \micron\ is important to break this degeneracy.
The best-fit $T_{\rm d}$ spans 80~K to 40~K and is lower for a larger $\beta$.
The best-fit $\log_{10}(M_{\rm d}/{\rm M}_\odot)$ increases from 6.6 to 7.5 with increasing $\beta$.
The corresponding total IR luminosity changes from $\log_{10}(L_{\rm IR}/{\rm L}_\odot)=12.0$ to \(11.6\).
The obtained values and their uncertainties are listed in appendix Table~\ref{tab:MdTdsummary}.
Although we find a smallest $\chi^2$ value for $\beta=1.0$, $\chi^2$ differences compared to $\beta=1.5$ or $2.0$ are not statistically significant.

\subsection{Radiative Equilibrium Fitting}
\label{sec:radeqfit}

\citet{Inoue.A:2020a} presented a new algorithm to derive dust temperature and mass using radiative equilibrium on dust grains.
Radiative equilibrium connects $T_{\rm d}$ with $M_{\rm d}$ and breaks degeneracy between them.
Therefore, we may obtain tighter constraints on them even without observing the FIR SED peak.
The algorithm requires the observed UV luminosity, $L_{\rm UV}$, and the physical radius of the system, $R$.
We adopt values taken from \citet{Hashimoto.T:2019a}.
The total UV luminosity is $L_{\rm UV}=(7.6\pm1.3)\times10^{44}$ erg~s$^{-1}$.
The observed full-width half-maximum along the major and minor axes of the entire dust emission in Band~6 are $a=3.8\pm1.1$ kpc and $b=0.8\pm0.5$ kpc in the proper coordinate, respectively.
The Band~6 observation in Hashimoto et al.\! has higher spatial resolution than our Band~7 observation.
Assuming a spherical symmetric structure for the analytic treatment of \citet{Inoue.A:2020a}, we adopt the radius of $R=\sqrt{ab}/2=0.87\pm0.30$ kpc.

\begin{figure}[t]
    \epsscale{1.0}
    \plotone{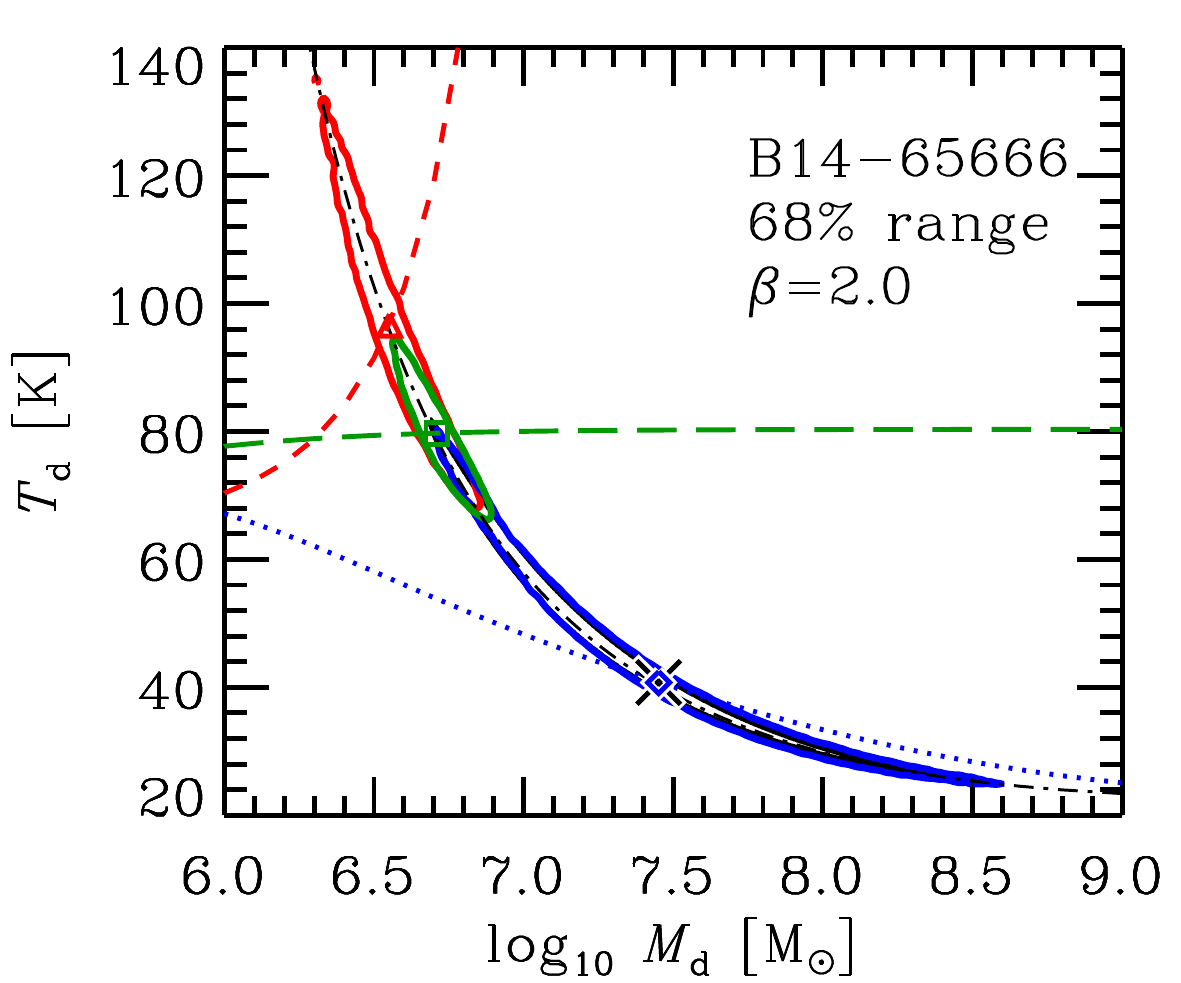}
    \caption{
    Best-fit solutions (symbols) and 68 percent areas (solid lines) in the radiative equilibrium fitting, and the dust temperature and mass plane for the emissivity index $\beta=2.0$. The red triangle, green square, and blue diamond represent the best-fit solutions in the shell, homogeneous, and clumpy geometries, respectively. The black cross represent the best-fit solution in the modified blackbody case. The dot-dashed line shows the dust temperature--mass relation of the observed Band~7 flux density. The red short-dashed, green long-dashed, and blue dotted lines show the dust temperature--mass relations in the radiative equilibrium for the shell, homogeneous, and clumpy models, respectively.
    }
    \label{fig:radeq_MdTd}
\end{figure}

\begin{figure}[t]
    \epsscale{0.9}
    \plotone{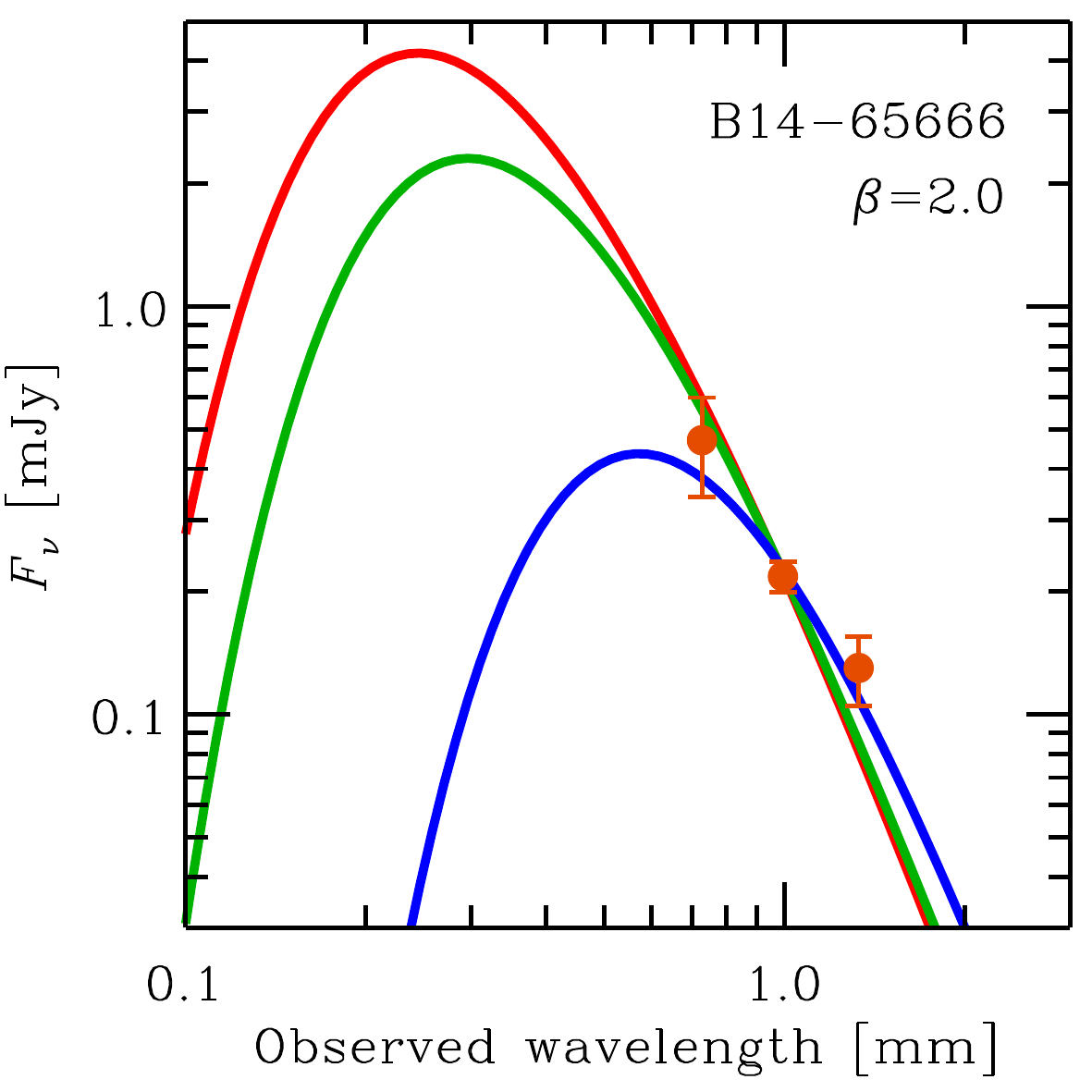}
    \caption{
    Best-fit FIR SEDs for the shell (red), homogeneous (green), and clumpy (blue) models. The data points with error bars are ALMA measurements. The emissivity index is $\beta=2.0$.
    }
    \label{fig:radeq_SED}
\end{figure}

\begin{figure}[t]
    \epsscale{0.9}
    \plotone{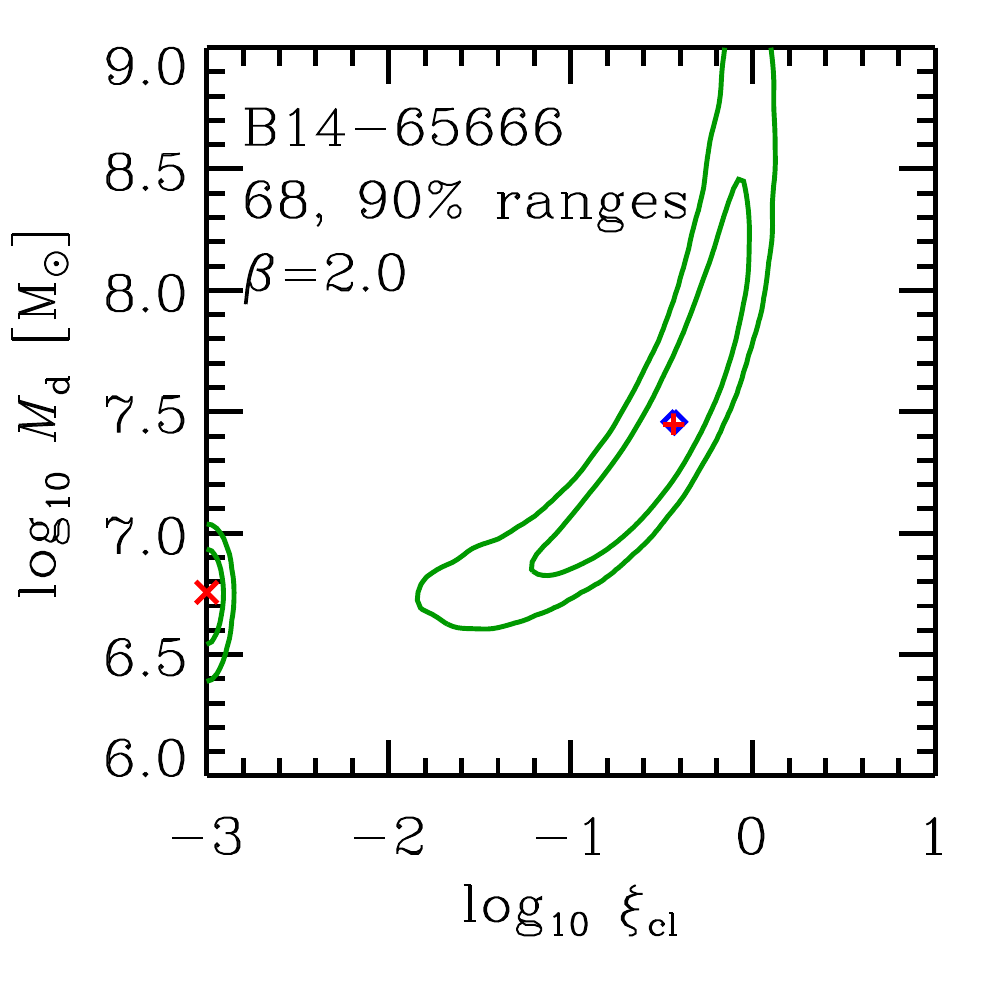}
    \caption{
    Distribution of the best-fit solutions obtained from 30,000 Monte Carlo trials for perturbed observational data in the clumpy model for the emissivity index $\beta=2.0$. The vertical and horizontal axes are dust mass and the clumpiness parameter, respectively. The contours enclose 68 percent and 90 percent of the solutions. The diamond shows the best-fit solution for the actual observational data. The plus sign and cross symbol represent the highest and second highest peaks of the density of the solutions, respectively. 
    }
    \label{fig:radeq_Mdclp}
\end{figure}

Following \citet{Inoue.A:2020a}, we perform least-$\chi^2$ fitting for FIR SED with the radiative equilibrium algorithm in three geometries: spherical shell, homogeneous sphere, and clumpy sphere.
The spherical shell and homogeneous sphere geometries require only a single free parameter, $M_{\rm d}$ (or $T_{\rm d}$), and the other quantity of $T_{\rm d}$ (or $M_{\rm d}$) is derived from $M_{\rm d}$ (or $T_{\rm d})$ thanks to the radiative equilibrium.
The clumpy geometry requires an additional free parameter to control the clumpiness, $\xi_{\rm cl}$, which is a non-dimensional parameter defined by the ratio between a single clump size relative to the entire system size and the volume filling factor of the clumps.
Figure~\ref{fig:radeq_MdTd} shows the fitting results for the case of $\beta=2.0$.
The other two $\beta$ cases are shown in appendix Figure~\ref{fig:radeq_MdTd_ax}.
Radiative equilibrium requires the relations between $T_{\rm d}$ and $M_{\rm d}$ shown by the short-dashed, long-dashed, and dotted lines for the shell, homogeneous, and clumpy geometries, respectively.
We consider the uncertainties of $L_{\rm UV}$ and $R$ in addition to FIR flux densities in the fitting using a Monte Carlo method\footnote{In each trial of the fitting, we fluctuate the observables assuming a Gaussian function with the standard deviation equal to the observed uncertainty. We then repeat the trials, and calculate the 68-percentile of the distribution of the best-fit values.} \citep{Inoue.A:2020a}.
Therefore, the resultant uncertainties in $M_{\rm d}$ and $T_{\rm d}$ are still large.
The clumpy geometry case gives the same best-fit solutions as those of the modified blackbody fitting because the clumpiness parameter functions as an adjuster \citep{Inoue.A:2020a}.
Figure~\ref{fig:radeq_SED} shows the best-fit FIR SEDs for $\beta=2.0$ and Figure~\ref{fig:radeq_Mdclp} shows the distribution of the solutions in the $M_{\rm d}$ and $\xi_{\rm cl}$ plane for $\beta=2.0$.
Other $\beta$ cases are found in Figures~\ref{fig:radeq_SED_ax} and \ref{fig:radeq_Mdclp_ax}, respectively.
The numerical values are summarized in appendix Table~\ref{tab:MdTdsummary}.
For the shell and homogeneous geometries, the cases of $\beta=1.5$ and 2.0 result in larger $\chi^2$ values and are not favored statistically compared to the case of $\beta=1.0$.
For the clumpy geometry, all $\beta$ cases cannot be regarded to be different statistically.

The best-fit $T_{\rm d}$ values are $\simeq100$~K in the shell geometry, $80$--100~K in the homogeneous geometry, and $40$--80~K in the clumpy geometry.
The corresponding $\log_{10}(M_{\rm d}/{\rm M}_\odot)$ values are \( 6.4\nn6.6 \), \( 6.5\nn6.7 \), and \( 6.6\nn7.5 \), respectively.
The IR luminosities are \( \log_{10}(L_{\rm IR}/{\rm L}_\odot)=12.5\nn13.0 \), \( 12.4\nn12.6 \), and \( 11.6\nn12.1\), respectively.
Comparing our results for B14-65666 to those for another $z\simeq7$ DSFG, A1689zD1 \citep{Inoue.A:2020a}, we find some similarities in the dust properties of those high-redshift objects.
In the shell and homogeneous geometries, both objects exhibit high IR luminosities and corresponding high SFRs, possibly indicating the invalidity of these simple geometries.
Hereafter, we adopt the case of the clumpy geometry at \( \beta=2.0 \) for the IR luminosity as the fiducial case.

In the clumpy geometry, the best-fit clumpiness parameter is \( \xi_{\rm cl}=0.1\nn0.4 \) for B14-65666, which is similar to that for A1689zD1.
As discussed in \citet{Inoue.A:2020a}, if a clump size is similar to the size of giant molecular clouds of $\sim10$ pc in the local Universe \citep{Larson.R:1981a, Fukui.Y:2008a}, these values correspond to a clump volume filling factor of 3\%--10\%.
Although observing such tiny clouds are difficult without significant gravitational lensing, a comparison between the clumpiness expected from the radiative equilibrium algorithm and galaxy formation simulations will be interesting.

\section{Discussion}
\label{sec:discussion}

\begin{figure}[t]
    \epsscale{1.2}
    \plotone{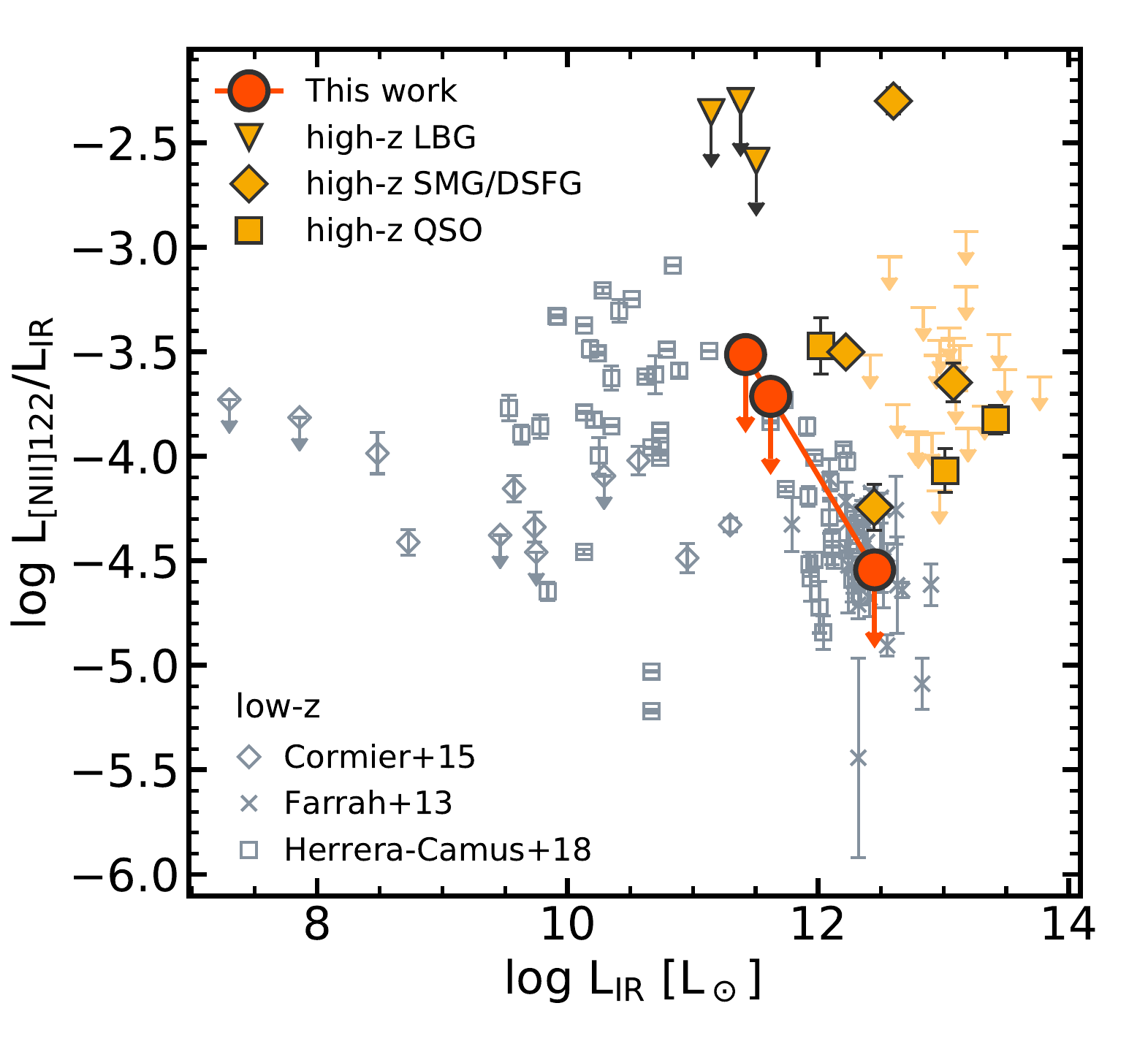}
    \caption{
      The ratio of [NII] 122 \um\ to the IR luminosity as a function of the IR luminosity.
      The red circles represent the constraints for B14-65666.
      The left and right red circles reflect the errors of the IR luminosity estimated with the radiative equilibrium fitting assuming the clumpy geometry at \( \beta = 2.0 \).
      The orange filled symbols show the measurements for high-redshift objects: three LBGs at \( z\sim6 \) \citep[upside-down triangles,][]{Harikane.Y:2020b}, SMGs/DSFGs at \( z=2\nn5 \) \citep[diamonds,][]{Ferkinhoff.C:2015a, Zhang.Z:2018a, Lee.M:2019a, De-Breuck.C:2019a}, and \( z>2 \) quasars \citep[squares,][]{Ferkinhoff.C:2015a, Novak.M:2019a, Lee.M:2019a, Li.J:2020b}, while the orange bars with upper limits show undetected DSFGs at \(z = 1\nn3.6\) \citep{Zhang.Z:2018a}.
      The gray open symbols are taken from studies on local galaxies: \citet[][diamonds]{Cormier.D:2015a}, \citet[][crosses]{Farrah.D:2013a}, and \citet[][squares]{Herrera-Camus.R:2018a}.
      In this figure, the error bars of the references only include the errors of the [\NII] luminosity.
      The upper limits for B14-65666 are located in the distribution of the local galaxies and do not show any excess as some of the high-redshift objects.
    }
    \label{fig:40}
\end{figure}

\subsection{Ratio of [\NII] to IR luminosity}
\label{sec:ratio-nii-ir}
In Figure \ref{fig:40}, we compare the ratio of the [\NII] 122 \um\ to IR luminosity (\(\LNii/L_\text{IR}\)) with the ratios in the literature.
The observed line-to-IR luminosity ratio decreases with increasing IR luminosity in local luminous infrared galaxies \citep[line deficit; e.g.,][]{Herrera-Camus.R:2018b, Herrera-Camus.R:2018a}.
Figure \ref{fig:40} shows \(\LNii/L_\text{IR}\) as a function of \(L_\text{IR}\).
The three red circles show the measurements of this study, reflecting the uncertainties of the IR luminosity in the clumpy geometry.
The gray open symbols depict the local reference measurements: dwarf galaxies \citep{Madden.S:2013a, Cormier.D:2015a}, ultra/luminous infrared galaxies \citep[][]{Farrah.D:2013a}, and star-forming, Seyfert, and luminous infrared galaxies \citep[][]{Herrera-Camus.R:2018a}.
The \(\LNii/L_\text{IR}\) upper limits of B14-65666 is on the relation of the local galaxies.
At high redshift, SPT 0418-47 exhibits an [\NII] luminosity consistent with the local galaxies \citep{De-Breuck.C:2019a}.
In contrast, some high-redshift objects are located above the local relations, like SMMJ02399-0136 \citep{Ferkinhoff.C:2011a, Ferkinhoff.C:2015a}, the Cosmic Eyelash \citep{George.R:2014a, Zhang.Z:2018a}, BRI 1202-0725 SMG \citep{Iono.D:2006a, Lee.M:2019a} and \( z>2 \) quasars \citep{Ferkinhoff.C:2015a, Novak.M:2019a, Lee.M:2019a, Li.J:2020b} as discussed in \citet{Li.J:2020b}.
Our analysis demonstrates that B14-65666 does not show an excess in the [\NII]-to-IR luminosity ratio from the local decreasing trend.
Compared with the upper limits of \(z = 1\nn3.6\) lensed DSFGs \citep{Zhang.Z:2018a} and \(z\sim6\) LBGs \citep{Harikane.Y:2020b}, our observation gave stringent upper limits of \(\LNii/L_\text{IR}\) for B14-65666 at a fixed \( L_\text{IR} \).

\begin{figure}[thbp]
    \epsscale{1.1}
    \plotone{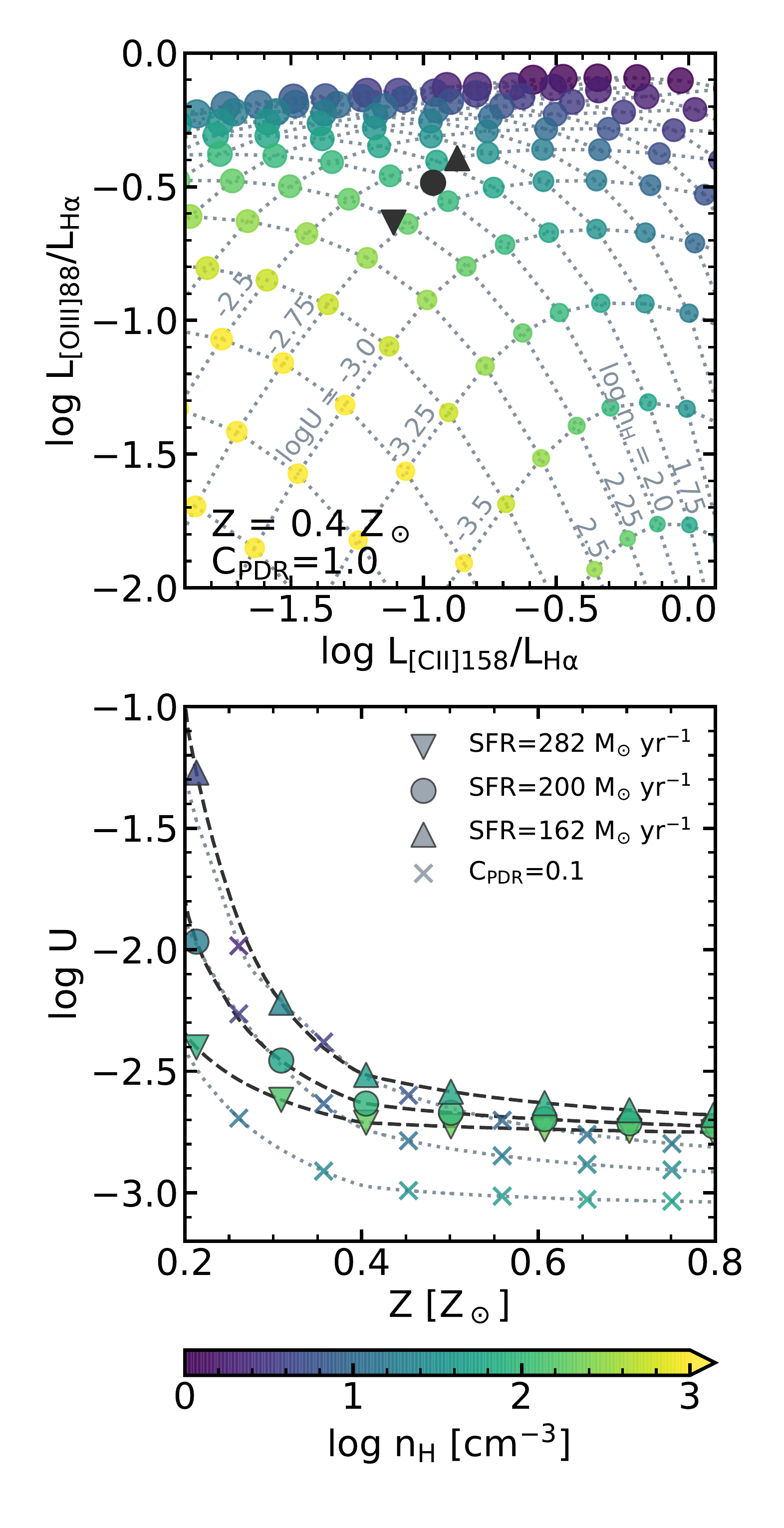}
    \caption{
      Top: \(\LOiii/\LHa\)--\(\LCii/\LHa\) diagram.
      The grid shows the result of the Cloudy model at the metallicity of \( 0.4 \) \Zo, where the circle sizes and colors depend on the ionization parameter \( U \) and hydrogen density \( n_\text{H} \), respectively.
      The ranges of parameters are \( -4.0 < \log_{10} U < -0.5 \) and \( 0 < \log_{10} n_\text{H}/\text{cm}^{-3} < 4.0 \) in steps of \( 0.25 \).
      The black symbols show the location of B14-65666 on the diagram, where \LHa\ is converted from the SED \SFR.
      The triangle, circle, and upside-down triangle show the \SFR\ differences of \( 162 \), \( 200 \), and \( 282 \) \Moyr, respectively.
      Bottom: Ionization parameter \( U \) as a function of the metallicity \( Z \) color-coded by hydrogen density \( n_\text{H} \).
      The three dashed sequences represent the differences of the \SFR.
      The dotted lines with the crosses represent the case of \( C_\text{PDR} = 0.1 \).
      At \( Z > 0.4 \) \Zo\ \( U \) is almost constant at \( \log_{10} U \simeq -2.7\pm0.1 \)  and \( n_\text{H} \) is \( 50\nn 250\) \cmmm, whereas at \( Z < 0.4 \) \Zo\ both nebular parameters drastically change.
    }
    \label{fig:50}
\end{figure}

\subsection{Estimates of Nebular Physical Parameters}
\label{sec:estim-phys-param}
Emission-line ratios originating from star-forming activities are determined by physical properties of nebulae around massive stars.
Given nebular parameters and ionizing radiation sources, photoionization models of the nebulae can be constructed, and the emission-line fluxes from various atoms and ions can be predicted.
In this section, we use a photoionization code Cloudy \citep{Ferland:1998a} to examine principal nebular parameters, i.e., the metallicity \(Z\), ionization parameter \(U\), and hydrogen density \(n_\text{H}\) at a surface illuminated by an ionizing radiation source.

The luminosities of [\OIII] 88 \um\ (\LOiii) and [\CII] 158 \um\ (\LCii) lines provide information on the nebular parameters.
We use the \(\LOiii/\LHa\)--\(\LCii/\LHa\) diagram, similar to the \(\LOiii/\SFR\)--\(\LCii/\SFR\)\ diagram proposed by \citet{Harikane.Y:2020b}, to estimate the nebular parameters.
Harikane et al.\ illustrated a diagram that compared \(\LOiii/\SFR\)\ and \(\LCii/\SFR\)\ as functions of \( U \), \( n_\text{H} \), \( Z \), and other parameters.
We adopt the concepts of Harikane et al.\ and \citet{Nagao.T:2011a, Nagao.T:2012a} to model \HII\ regions and PDRs in a plain-parallel geometry using a software Cloudy version 17.01 \citep{Ferland.G:2017a} under the assumptions of the pressure equilibrium, an identical metallicity in the stellar and gas phases, and the solar \(\text{C} / \text{O}\) abundance.
Following \citet{Inoue.A:2014a}, in our model, we use input spectral shapes of 10-Myr constant star-formation models at stellar metallicities of \( Z_{*}= 0.02, 0.2, 0.4, \) and \( 1.0 \) \Zo, created with the \textsc{Starburst99} \citep{Leitherer:1999a} with a Salpeter initial mass function at \( 1\nn100\) \Mo.
We also test other input spectral shapes created with the Binary Population and Spectral Synthesis code \citep[BPASS,][]{Eldridge.J:2017a} version 2.2.1 \citep{Stanway.E:2018a} to find that the results are qualitatively the same and that our conclusions do not change.
Parameter grids of \( U \) and \( n_\text{H} \) are \( -4.0 < \log_{10} U < -0.5 \) and \( 0 < \log_{10} n_\text{H}/\text{cm}^{-3} < 4.0 \) in steps of \( 0.25 \) dex.
The software is run until the \textit{V}-band dust extinction reaches 100 mag \citep{Abel.N:2005a}.
Cloudy outputs emission line strengths relative to an \Hb\ line.
From [\OIII] 88 \um\(/\)\Hb, [\CII] 158 \um\(/\)\Hb, and \Ha\(/\)\Hb\ line ratios, we computed [\OIII]\(/\)\Ha\ and [\CII]\(/\)\Ha\ line ratios, that is, \(\LOiii/\LHa\) and \(\LCii/\LHa\) luminosity ratios.

The free parameters in our model are \( U \), \( n_\text{H} \), \( Z \), and the PDR covering fraction \CPDR.
\CPDR\ is a posterior parameter described in \citet{Cormier.D:2019a} and \citet{Harikane.Y:2020b}, regulating line intensities emitted from the PDR.
Geometrically, it means what fraction of a surface of an \HII\ region is covered by a PDR: the \HII\ region is entirely covered by the PDR when \(\CPDR = 1\), while a part of the \HII\ region is not covered for a low \CPDR\ case and thus [\CII] line intensities from the PDR are scaled by a factor of \(\CPDR\).
When \( Z \) and \CPDR\ are fixed, a nebular parameter pair (\( U \), \( n_\text{H} \)) is in one-to-one correspondence to (\(\LOiii/\LHa\), \(\LCii/\LHa\)), as shown in the top panel of Figure \ref{fig:50}.
We derive \( U \) and \( n_\text{H} \) as functions of \( Z \), for \( \CPDR = 1.0 \) and \( 0.1 \) as fiducial parameters.

The \SFR\ and gas-phase metallicity of B14-65666 are almost constrained to be \( \SFR = 200^{+82}_{-38} \) \Moyr\ and \( Z= 0.4^{+0.4}_{-0.2} \) \Zo\ using SED fitting in \citet{Hashimoto.T:2019a}.
SED fitting uses photometry from NIR to ALMA FIR bands (from UV to FIR bands in the rest frame), including the effect of dust attenuation.
The estimated metallicity is consistent with another metallicity estimates from the [\OIII] 88 \um\ luminosity and \SFR\ by \citet{Jones.T:2020a}.
Compared to the errors of \( \simeq 12\% \) in \LOiii\ and \LCii, the \SFR\ error is large and dominates measurement uncertainties in the results.
We, therefore, evaluate uncertainties in the nebular parameters using \( \SFR = 162 \), \( 200 \), and \( 282 \) \Moyr\ at metallicities from \( Z=0.2 \) to \( 0. 8\) \Zo.
We obtain \(\LOiii/\LHa\) and \(\LCii/\LHa\) ratios for B14-65666 by computing \LHa\ from the SED \SFR\ using Equation 2 in \citet{Kennicutt.R:1998a} with a correction factor of \(0.63\) \citep{Madau:2014} from \citet{Salpeter:1955} to \citet{Chabrier:2003} initial mass functions.
At each metallicity, (\(\LOiii/\LHa\), \(\LCii/\LHa\)) values can be converted into (\( U \), \( n_\text{H} \)) values through linear interpolation.
Figure \ref{fig:50} displays the \(\LOiii/\LHa\)--\(\LCii/\LHa\)\ diagram at \( Z=0.4 \) \Zo\ and \( \CPDR=1.0 \) in the top panel.
The model grid in the diagram depends on the \( Z \) and \CPDR\ values.
The modeled \LOiii\(/\)\LHa values increase with an increase in metallicity due to high oxygen abundances, whereas the modeled \LCii\(/\)\LHa values decrease with a decrease in \CPDR\ by almost the same factor.
We refer readers to \citet{Harikane.Y:2020b} for detailed characteristics of the models.

The bottom panel of Figure \ref{fig:50} shows the estimated \(U\) values as a function of \(Z\), color-coded by \( n_\text{H} \).
The triangles, circles, and upside-down triangles show the low (\( 162 \)), middle (\( 200 \)), and high (\( 282 \) \Moyr) \SFR\ cases for \( \CPDR=1.0 \), respectively.
Both \( U \) and \( n_\text{H} \) monotonically decreases and increases at a fixed \SFR, respectively, as the assumed metallicity increases.
At high metallicities of \( Z > 0.4 \) \Zo, the ionization parameter is almost constant at \( \log_{10} U \simeq -2.7\pm0.1 \) and the hydrogen density is \( n_\text{H} \simeq 50\nn 250\) \cmmm.
The dispersion of \( U \) caused by \SFR\ uncertainty is small.
The \( U \) values are comparable with those in local dwarf \citep{Cormier.D:2019a} and \( z\sim2 \) galaxies \citep{Strom.A:2018a} in similar metallicity range, as well as nearby starburst galaxies with higher metallicity \citep{Herrera-Camus.R:2018b}.
Specifically, these \( U \) values are on \( U\nn Z \) invariant relation from local to \( z\sim2 \) galaxies \citep{Sanders.R:2020a}.
The range of the hydrogen density of our galaxy is between the average values of that of local and \( z\sim2 \) galaxies \citep{Sanders.R:2016a}.
We do not find any further increase of hydrogen density from \( z\sim2 \) to \( 7 \) even though hydrogen density increases with an increase in redshift by a factor of 10 from \( z\sim0 \) to \( 2 \) galaxies \citep[e.g.,][]{Sanders.R:2016a}.
In contrast, at \( Z < 0. 4 \) \Zo, \( \log_{10} U \) drastically increases up to \( -1.0 \) with a decrease in metallicity.
Hydrogen density simultaneously drops to \( n_\text{H} \sim 1 \) \cmmm.
Such extreme conditions of the nebular parameters are hardly found in normal galaxies at low redshifts, whereas dwarf galaxies with low metallicity and low specific SFR tend to exhibit similar high \( U \) and low \( n_\text{H} \) \citep{Cormier.D:2019a}. 
These drastic changes in the nebular parameters are caused by the high \(\LOiii/\LHa\)\ ratio (i.e., high \(\LOiii/\SFR\)) of B14-65666, which is difficult to explain in low metallicity regimes in our model.
Notably, our model cannot reproduce the high \(\LOiii/\LHa\)\ ratio at \( Z < 0.1 \) \Zo\ even in the high \SFR\ case.
The crosses in Figure \ref{fig:50} depict the models at \(\CPDR = 0.1\).
In this case, \( U \) and \( n_\text{H} \) become smaller than the \( \CPDR = 1.0\) case at a fixed metallicity.
Specifically, \( n_\text{H} \) is predicted to be low; the values are smaller by an order of magnitude at \( Z > 0.4 \) \Zo\ and \( n_\text{H} \lesssim 10 \) \cmmm\ at \( Z < 0.4 \) \Zo.
However, qualitative tendencies of \( U \) and \(n_\text{H} \) to \( Z \) are unaffected by \CPDR\ difference.

It is informative to compare our photoionization models with previous works focusing on [\OIII] and [\NII] lines.
\citet{Rigopoulou.D:2018a} used the ionization models in \citet{Pereira-Santaella.M:2017a} to estimate gaseous metallicity from a line ratio of [\OIII] 88 \um\ to [\NII] 122 \um\ ([\OIII]\(/\)[\NII]), under the assumption of a local \NO--\(Z\) relation.
The [\OIII]\(/\)[\NII] ratio is insensitive to hydrogen density but sensitive to the metallicity and ionization parameter \citep{Pereira-Santaella.M:2017a}.
Rigopoulou~et~al.\ firstly inferred the ionization parameter from 88-to-122 \um\ dust continuum ratios, and then derived the metallicity from the ionization parameter and [\OIII]\(/\)[\NII] ratio.
The dust continuum ratio of B14-65666 is \(\simeq\! 1.4\nn2.9 \) (Table \ref{tab:1}), corresponding to \( \log_{10} U\sim -2\) to \(-1\) at \( n_\text{H} = 100 \) \cmmm, which is higher than our model predictions in average.
Our measurement of [\OIII]\(/\)[\NII] \( \gtrsim 40 \) for B14-65666 gives a weak metallicity constraint of \( Z \lesssim 1\) \Zo\ for \( \log_{10} U \sim -2 \) (the highest ionization parameter value in Pereira-Santaella~et~al.\! model) and a marginal constraint of \( Z \lesssim 0.6 \) \Zo\ for \( \log_{10} U \simeq -2.7 \), which is consistent with the metallicity range estimated from the SED fitting \citep{Hashimoto.T:2019a}.
The reason of the high ionization parameter inferred from the 88-to-122 \um\ dust continuum ratio is unclear, but it may be related with the fact that high-redshift galaxies tend to exhibit higher dust temperatures \citep{Bakx.T:2020a} than local galaxies that were used for their model calibration.

Although we assume that [\CII] is emitted from \HII\ regions and PDRs, [\CII] emission also arises from shock excitation caused by galaxy interactions.
It can be a cause for concern that shocks contribute to the [\CII] line flux since B14-65666 exhibits a merger morphology \citep{Hashimoto.T:2019a}.
\citet{Appleton.P:2013a} measured the [\CII] flux originating from shocks by observing the intergalactic medium in Stephan’s Quintet.
They report that the [\CII]-to-IR luminosity ratios are \( > 10^{-1.5} \), which is higher than those measured in star-forming galaxies, including B14-65666 \citep[\( \sim10^{-2.5} \),][]{Hashimoto.T:2019a}.
The [\CII] luminosity by shocks (\( \sim10^7 \) \Lo) is also lower than for B14-65666 (\( \sim 10^9 \) \Lo).
We, therefore, conclude that shock excitation is less dominant in [\CII] emission in B14-65666 unless the shock mechanism significantly differs between the two objects.

\begin{figure*}[t]
    \epsscale{1.2}
    \plotone{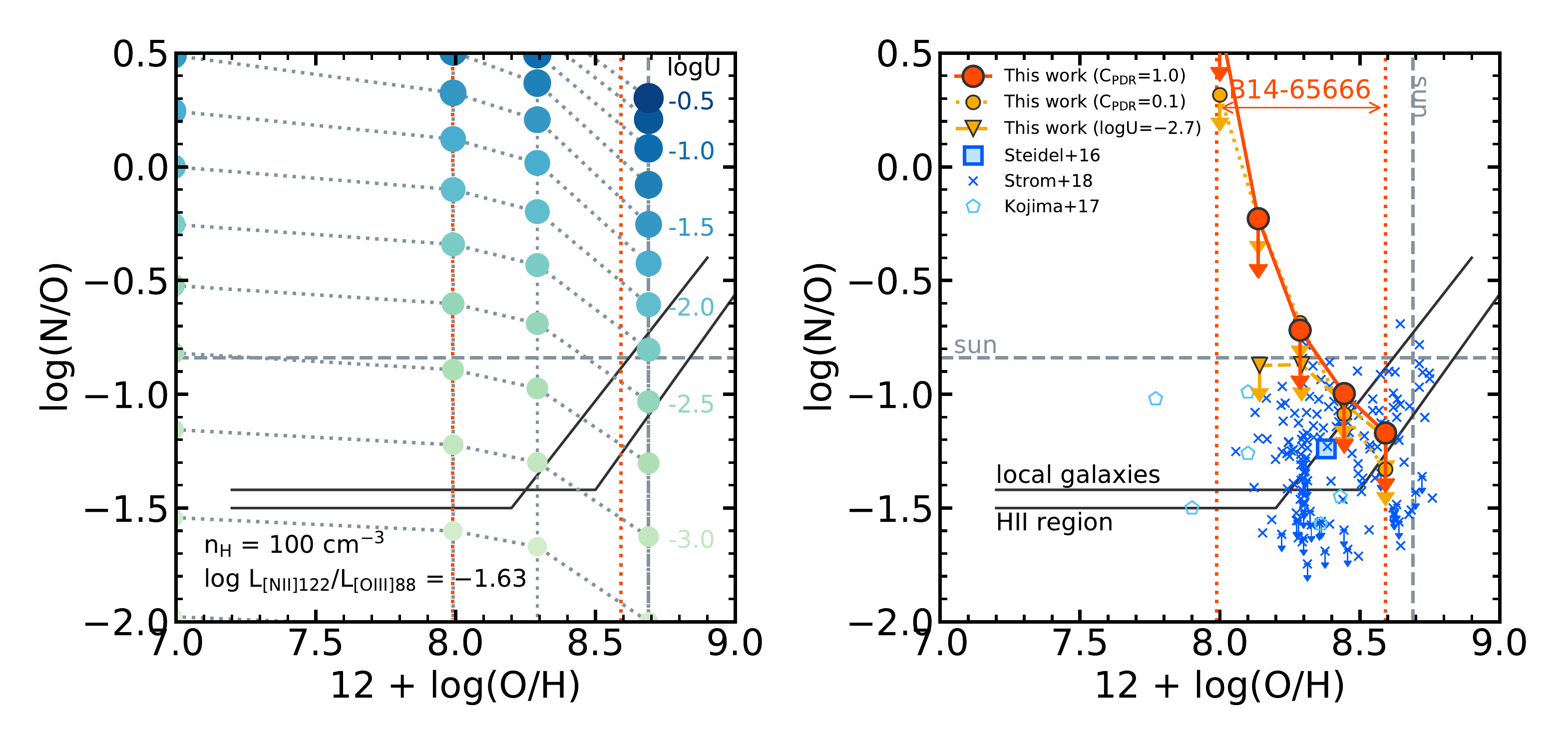}
    \caption{
      The nitrogen-to-oxygen abundance ratio as a function of the metallicity.
      Left: Cloudy model grid as a function of metallicity and ionization parameter, in the case of \( \log n_\text{H} = 2.0 \) and \( \log \LNiiLOiii = -1.63 \), which is the observed \( 3\sigma \) upper limit of a luminosity ratio.
      The symbol colors and sizes depend on the ionization parameter.
      The black solid lines show the average relations of extragalactic HII regions \citep{Pilyugin.L:2012a} and local SDSS galaxies \citep{Andrews.B:2013a} estimated using the direct temperature method.
      The gray dashed lines indicate the solar values.
      Right: The red circles show the \NO\ upper limits of B14-65666 at \( C_\text{PDR}=1.0 \) as a function of  metallicity, obtained from the measured luminosity ratio, \( \log \LNiiLOiii < -1.63 \).
      The range of the metallicity inferred from the SED fitting by \citet{Hashimoto.T:2019a} is \( 12+\logOH = 7.97\nn8.57\), corresponding to \( Z = 0.2\nn0.8\) \Zo\ (red dotted line).
      The small orange circles show the case of \( C_\text{PDR} = 0.1 \).
      The orange upside-down triangles show the case in which the ionization parameter is fixed at \( U = -2.7 \).
      The blue and cyan data points show the \( z\sim2 \) galaxies without error bars:
      \citet[][blue crosses]{Strom.A:2018a},
      \citet[][blue square]{Steidel:2016},
      and \citet[][cyan pentagon]{Kojima.T:2017a}.
      The \NO\ ratio is constrained to be sub-solar if the metallicity of B14-65666 is \( 12+\logOH \gtrsim 8.4 \).
    }
    \label{fig:60}
\end{figure*}

\subsection{Nitrogen-to-Oxygen Abundance Ratio}
\label{sec:no-abundance}
As seen in dwarf galaxies \citep[e.g.,][]{Lequeux.J:1979a,Vila-Costas.M:1993a} and SDSS galaxies \citep{Andrews.B:2013a}, the \NO\ abundance ratio is almost constant as \( \logNO \simeq -1.5 \) at low metallicity, and it drastically increases when metallicity exceeds a certain value.
A simple explanation of this \NO\ trend is a combination of the primary and secondary nucleosynthetic nitrogen; the primary production of nitrogen is independent of the initial metallicities in stars, whereas the secondary production that occurred in the CNO cycle is proportional to the initial carbon or oxygen abundance \citep[][]{Pagel.B:2009a}.

The \NO\ ratio affects the intensity ratios between nitrogen and oxygen atoms/ions emission lines.
In some photoionization models, the \NO\ ratio is assumed as a function of  metallicity \citep{Nagao.T:2011a, Pereira-Santaella.M:2017a, Rigopoulou.D:2018a}, but it is possible that this relation changes at high redshift.
Given the nebular parameters and input spectrum (i.e., a certain fixed ionization structure), photoionization models can predict abundance ratios from observed emission-line intensity ratios.
In FIR bands, [\NIII] 57 \um\(/\)[\OIII] 52 \um\ line ratio is used for \NO\ measurements \citep[e.g.,][]{Lester.D:1983a, Rubin.R:1988a, Peng.B:2021a}, as both lines have similar ionization potentials and critical densities.
For B14-65666, [\NIII] 57 \um\ and [\OIII] 52 \um\ lines are accessible in ALMA Band 9; however, they still requires expensive observations.
Instead, by assuming the nebular parameters derived from the [\OIII] and [\CII] emission lines (Section \ref{sec:estim-phys-param}), we constrain the \NO\ abundance ratio at \( z\sim7 \) from our upper limits of the luminosity ratio between [\NII] 122 \um\ and [\OIII] 88 \um\ (\LNiiLOiii).

We convert \LNiiLOiii\ to \NO\ with Cloudy models and the nebular parameters obtained in Section \ref{sec:estim-phys-param}, which are functions of metallicity (see Figure \ref{fig:50}).
The observed line luminosity ratio is \(\LNiiLOiii < -1.63\).
For each nebular parameter set, we prepare models with \logNO\ ranging from \( -2 \) to \( 0 \) in steps of \( 0.5 \).
We note that \( n_\text{H} \) is included in our calculations even though \( \LNii/\LOiii \) is almost independent of \( n_\text{H} \) due to similar critical densities between [\NII] 122 \um\ and [\OIII] 88 \um.
We compare the observed \( \LNii/\LOiii \) ratios and the Cloudy models to obtain the upper limits of \NO\ as a function of \( U \) and \( Z \).
The left panel of Figure \ref{fig:60} shows a Cloudy model grid in the case of \( \log n_\text{H} = 2.0 \) and \( \log \LNiiLOiii = -1.63 \).
At a fixed \( Z \), higher \( U \) results in a higher \NO\ upper limit (i.e., a weaker \NO\ constraint).
Since the low \SFR\ yields high-\(U\) solutions among the three \SFR\ cases considered in Section \ref{sec:estim-phys-param}, the estimated \NO\ upper limits become the highest (weakest) at the low \SFR.
To simply express \NO\ as a function of \(Z\), we take the highest \NO\ ratios as a upper limit at each \( Z \) in the following.

The right panel of Figure \ref{fig:60} illustrates the upper limits of the \NO\ abundance ratio for B14-65666 as a function of metallicity.
Given that the solar value of \( 12+\logOH \) is \( 8.69 \) \citep{Asplund.M:2009a}, the metallicity \( Z \) is converted to the \OH\ abundance ratio.
The \NO\ value is well constrained at a high metallicity regime.
This originates from a model prediction of relatively high \LNiiLOiii\ for the range of ionization parameters at high metallicities.
At \( 12+\logOH \gtrsim 8.4 \), \logNO\ is less than the solar value of \( -0.84 \) \citep{Asplund.M:2009a}.
At a higher metallicity of \( 12+\logOH > 8.5 \), the \logNO\ upper limits become less than the average relation of extragalactic \HII\ regions \citep{Pilyugin.L:2012a}.
This is consistent with an implication by the \citet{Rigopoulou.D:2018a} model that the metallicity of B14-65666 is \( \lesssim 0.6 \) \Zo\ if \( \log_{10} U \simeq - 2.7 \) and the \NO\ ratio follows a local \NO--\(Z\) relation (Section \ref{sec:estim-phys-param}).
On the other hand, the \NO\ constraint is very weak at \( 12+\logOH < 8.3 \) due to high ionization parameters of \( \log U \gtrsim -2.5 \).
If the metallicity of B14-65666 is \( 12+\logOH < 8.3 \), our observations cannot produce a meaningful constraint on the \NO\ abundance, implying that we need more sensitive observations by 1--2 order of magnitude to detect the [\NII] 122 \um\ emission line from the \( z\sim7 \) galaxy.
The small orange circles in Figure \ref{fig:60} depict the case of \( \CPDR=0.1 \).
We find that the low PDR covering fraction only weakly affects the upper limits of the \NO\ ratio by 0.3-dex at most.
As shown in Figure \ref{fig:50}, the low \CPDR\ slightly changes \( U \) in the low \SFR\ case, which influences the \NO\ upper limit.
Although \( n_\text{H} \) decreases in this case, \LNiiLOiii\ is almost independent of \( n_\text{H} \) and the N/O constraints are not affected.
Therefore, the lower \CPDR\ does not change the \NO\ abundance very much.

The weak \NO\ upper limits at \( 12+\logOH < 8.3 \) originates from the high ionization parameters of \( \log_{10} U \gtrsim -2.5 \).
If the ionization parameter of B14-65666 is similar to those of local and \( z\sim2 \) galaxies, the upper limits would become more stringent.
We test whether this case is possible for our galaxy, by fixing \( \log_{10} U = -2.7 \) and by changing \CPDR\ from \( 0.05 \) to \( 1.0 \) on the \(\LOiii/\LHa\)--\( \LCii/\LHa \) diagram.
In this analysis we choose nebular parameter sets that can model the observed value within the \SFR\ uncertainty of \( 162<\SFR<282 \) \Moyr\ and take the highest \NO\ in the parameter sets  as the upper limits at fixed metallicities.
The results are depicted with the orange upside-down triangles in Figure \ref{fig:50}.
The assumption of \( \log_{10} U = -2.7 \) gives similar upper limits to the \( \CPDR=1.0 \) case (red circles) at \( 12+\logOH > 8.3 \), while the upper limit is constant at \( \logNO=-0.87 \) at lower metallicities of \( 12+\logOH < 8.3 \), as predicted from the left panel of Figure \ref{fig:60}.
This \NO\ upper limit is 3 times lower than those shown by the red and orange circles for which \( \log U \) increases as the metallicity decreases.
If \( \log_{10} U = -2.7 \), the \NO\ ratio of this galaxy is restricted to be sub-solar, irrespective of its intrinsic metallicity.
Our model requires a low PDR covering fraction (\( \CPDR < 1 \)) and high \SFR\ (\( 282 \) \Moyr) to keep \( \log_{10} U = -2.7 \) at lower metallicities.
We cannot plot the data point at \( 12+\logOH = 8.0 \) in the figure because there are no models with \( \log_{10} U = -2.7 \) at this metallicity within the \SFR\ uncertainty interval.

Our \NO\ upper limit is roughly consistent with \( z\sim2 \) studies.
\citet{Steidel:2016} stacked the KBSS-MOSFIRE spectra to compute the typical \NO\ of them and \citet{Strom.A:2018a} estimated \NO\ for individual KBSS-MOSFIRE galaxies with photoionization models.
These \NO\ ratios are comparable to those of the local extragalactic \HII\ regions presented by \citet{Pilyugin.L:2012a}.
The relatively low metallicity galaxies at \( z\sim2 \) studied by \citet{Kojima.T:2017a} exhibit higher \logNO\ than local galaxies at fixed metallicities, which are measured using a direct temperature method.
Most of these \NO\ ratios at \( z\sim2 \) are lower than the upper limit of B14-65666.
Our upper limits are also consistent with the nearby analogs of LBGs in \citet{Loaiza-Agudelo.M:2020a}.

\subsection{Predicted BPT Diagram}
\label{sec:bpt-diagram}
As noted in the previous section, the photoionization model can convert an emission-line flux into another line flux for the same ion.
This enables us to predict the position of B14-65666 on the BPT diagram from our FIR line fluxes.
We estimate optical-line ratios of \( [\OIII]\lambda5007/\Hb \) and \( [\NII]\lambda6583/\Hb \) from the FIR [\OIII] 88 \um\ and [\NII] 122 \um\ line fluxes, respectively, using the Cloudy models with the nebular parameters obtained in Section \ref{sec:estim-phys-param}.
In this model, \( [\NII]\lambda6583/\Hb \) is given as an upper limit because the [\NII] 122 \um\ flux is constrained as the \(3\sigma\) upper limit in our observations.
The \( \Ha/\Hb \) line ratio is also computed from the model.
In this way, we obtain the modeled \( [\OIII]\lambda5007/\Hb \) and \( [\NII]\lambda6583/\Ha \) line ratios for B14-65666.

\begin{figure}[t]
    \epsscale{1.2}
    \plotone{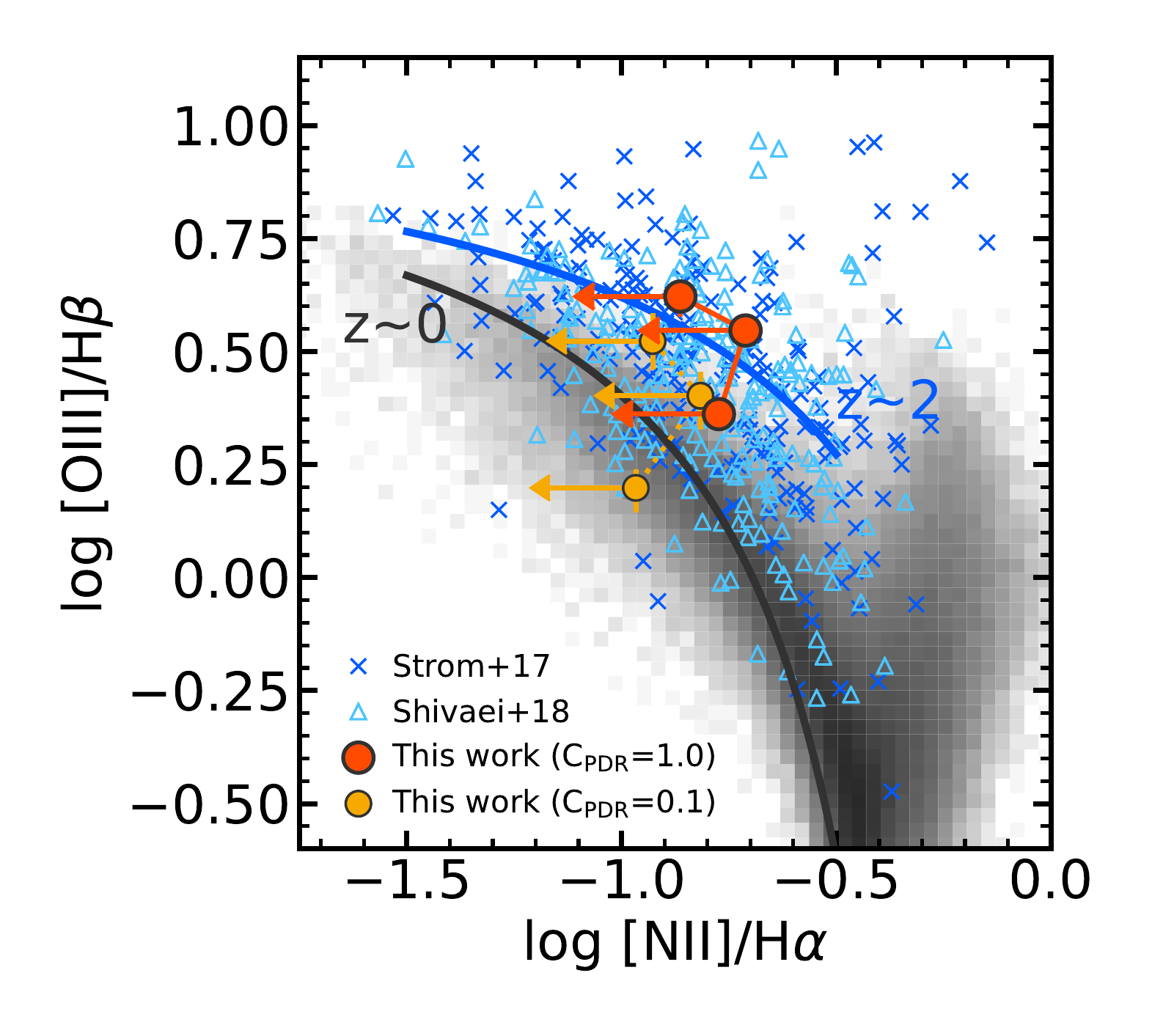}
    \caption{
      Estimated location of the B14-65666 upper limits on the BPT diagram.
      The location is computed from the photoionization model with the nebular parameters \( U \), \( n \), and \( Z \) (see the text in Section \ref{sec:bpt-diagram}).
      The red circles are assumed as \( C_\text{PDR}=1.0 \) and \( Z = 0.2, 0.4\), and \( 0.8  \) \Zo, where low metallicity data points show high [\OIII]\(/\)\Hb.
      The orange small circles denote the locations in the case of \( C_\text{PDR} = 0.1 \).
      The black density map indicates local SDSS galaxies.
      The blue crosses and cyan triangles indicate star-forming galaxies at \( z\sim2 \) from the KBSS-MOSFIRE survey \citep{Strom.A:2017b} and the MOSDEF survey \citep{Shivaei.I:2018a}, respectively.
      The black and blue lines represent the best-fit relations for the \( z\sim0 \) SDSS galaxies and the \( z\sim2 \) KBSS-MOSFIRE galaxies, respectively \citep{Strom.A:2017b}.
      As seen, B14-65666 is expected to located on or below the \( z\sim2 \) relation.
    }
    \label{fig:70}
\end{figure}

Figure \ref{fig:70} plots a BPT diagram.
In general, dust attenuation is negligible on the BPT diagram because the diagram takes the ratios of the emission lines with close wavelengths.
The FIR [\OIII] 88 \um\ and [\NII] 122 \um\ fluxes are not affected by dust attenuation due to their long wavelengths.
For these reasons, our estimations can be compared with values in literature without concerns about dust attenuation.

The estimated optical-line ratios for B14-65666 are \( \log_{10}([\OIII]/\Hb) = 0.55\pm0.02 \) and \( \log_{10}([\NII]/\Ha) < -0.71 \) at \( Z = 0.4 \) \Zo.
The \( [\OIII]/\Hb \) ratios become high at \( Z = 0.2 \) \Zo\ and low at \( Z=0.8 \) \Zo, while both \( [\NII]/\Ha \) upper limits become lower than the value at \( Z = 0.4 \) \Zo.
These values are depicted with the red circles in Figure \ref{fig:70}.
We note that B14-65666 is located in the region where star-forming galaxies are distributed on the BPT diagram \citep{Kauffmann.G:2003b}, which is a natural consequence of including the starbursting spectrum into the Cloudy model calculations (Sec \ref{sec:estim-phys-param}).
We are assuming that the entire emission-line fluxes originate from star formation not from AGNs.
When \( \CPDR = 0.1 \), the data points shown by the small orange circles move to the lower left, reflecting the low ionization parameter and hydrogen density.

To compare our results with low-redshift galaxies, we plot the distributions of local SDSS and \( z\sim2 \) star-forming galaxies in Figure \ref{fig:70}.
The flux ratios of SDSS galaxies are taken from the MPA/JHU catalog\footnote{URL: \url{https://wwwmpa.mpa-garching.mpg.de/SDSS/DR7/}}.
The \( z\sim2 \) galaxies are taken from the KBSS-MOSFIRE survey \citep{Strom.A:2017b} and the MOSDEF survey \citep{Shivaei.I:2018a}.
Our upper limits shown with the red circles (\( \CPDR=1.0 \)) are located above the local average relation and around the \( z\sim2 \) relation \citep{Strom.A:2017b}.
It is quite possible that B14-65666 is located below the average relation of the \( z\sim2 \) galaxies even though our model provides only upper limits, considering the weak [\NII] 122 \um\ flux suggested from the \NO\ upper limits at \( Z < 0.4 \) \Zo\ (Figure \ref{fig:60}).
In the case of \( \CPDR=0.1 \) (small orange circles), the upper limits are located around the \( z\sim0 \) galaxies.
At \( Z=0.8 \) \Zo\ in this case, especially low \( [\NII]/\Ha \) and \( [\OIII]/\Hb \) values are predicted because the [\NII] 122 \um\ line was undetected despite the low metallicity and low ionization parameters.
In summary, our model predicts that B14-65666, at \( z\sim7 \), is located on or below the average relation at \( z\sim2 \) on the BPT diagram.
If \CPDR\ is low, the location will become close to the \( z\sim0 \) relation.

If B14-65666 is representative of the entire galaxy population at \( z\sim7 \), we can discuss the redshift evolution from \( z\sim7 \) to \( 2 \) on the BPT diagram by comparing the location of B14-65666 with the \( z\sim2 \) relation.
However, the large dispersions of the galaxy distributions on the BPT diagram at \( z\sim0 \) and \( 2 \) raise a possibility that B14-65666 is not on the average relation at \( z\sim7 \).
In addition, the high UV luminosity and merger geometry of B14-65666 may not support the assumption that this galaxy is representative of the typical galaxies at \( z\sim7 \).
Further ALMA observations will explore the average ionization properties and distribution on the BPT diagram of the high-redshift galaxies.
More directly, the upcoming JWST will provide us opportunities to plot the BPT diagram at high-redshift by observing B14-65666 and other high-redshift galaxies in near- to mid-infrared bands.

\section{Summary}
\label{sec:conclusion}

We have performed ALMA Band 7 observations of a LBG at \( z=7.15 \), B14-65666, to target the [\NII] 122 \um\ FIR fine-structure line and underlying dust continuum emission.
B14-65666 is the first object detected in [\OIII] 88 \micron, [\CII] 158 \micron, and dust continuum emission at such high-redshift \citep[``Big Three Dragons'',][]{Hashimoto.T:2019a}.

The dust continuum at 120 \um\ is detected with \( \SN = 18.9 \).
We combine the dust-continuum flux at 120 \um\ with the previous measurements at 90 \um\ and 160 \um\ to perform two types of FIR SED fitting.
The modified blackbody fitting results in a dust temperature \( T_\text{d} = 80 \) to \( 40  \) K and a dust mass $M_\text{d} \sim 10^{6.6}$ to \( 10^{7.5} \) \Mo\ with an emissivity index \( \beta =1 \) to \( 2  \).
The corresponding IR luminosity spans \( \log_{10}(L_{\rm IR}/{\rm L}_\odot)=12.0 \) to \( 11.6 \).
The results of the radiative equilibrium fitting, proposed by \citet{Inoue.A:2020a}, are found to be similar to the results for another \( z\simeq7 \) dusty star-forming galaxy, A1689zD1.
Simple assumptions of the shell and homogeneous geometries appear to be invalid because the geometries prefer too high IR luminosity.
The clumpy geometry leads to the same best-fit results as the modified blackbody, with a best-fit clumpiness parameter of \( \xi_{\rm cl}=0.1\nn0.4 \).

The [\NII] 122 \um\ emission line is not detected.
The 3\( \sigma \) upper limit of [\NII] luminosity is \( 8.1 \times 10^7 \) \Lo.
We constrain the nebular parameters of B14-65666 as functions of metallicity with a photoionization code Cloudy, by modeling the [\NII] 122 \um\ upper limits, along with the [\OIII] 88 \um\ and [\CII] 158 \um\ line fluxes and the SED \SFR.
If the metallicity of B14-65666 is high (\( Z > 0.4\) \Zo), the ionization parameters and hydrogen densities are \( \log_{10} U \simeq -2.7\pm0.1\) and \( n_\text{H} \simeq 50\nn250 \) \cmmm, respectively.
The two nebular parameter values are consistent with those measured in low-redshift galaxies.
If \( Z < 0.4\) \Zo, the \( U \) and \( n \) drastically increases and decreases, respectively, with a decrease in metallicity.
This is due to the high \(\LOiii/\LHa\) ratio, that is, the observed high \(\LOiii/\SFR\)\ ratio of this galaxy.
In the case of a low PDR covering fraction (\(\CPDR = 0.1\)), lower \( U \) and \( n_\text{H} \) are expected, while the results are qualitatively the same.

The constraints on the nitrogen-to-oxygen abundance ratio, \NO, also largely depend on the assumed metallicity.
The obtained upper limit of the \NO\ ratio monotonically decreases as the assumed metallicity increases.
At \( 12+\logOH \gtrsim 8.4 \), the \NO\ ratios sould be sub-solar and upper limits are comparable to the \NO\ ratio of local and \( z\sim2 \) galaxies.
In contrast, our observations cannot provide meaningful constraint at \( 12+\logOH < 8.3 \).
The \NO\ ratio is insignificantly affected by the differences of the PDR covering fractions between \(\CPDR = 0.1\) and \( 1.0 \).
If we fix the ionization parameter at \( \log U = -2.7 \) in our model, the \NO\ ratios are restricted to be sub-solar even at \( 12+\logOH < 8.3 \) with a small PDR covering fraction and a high \SFR.

The Cloudy models also predict the location of the galaxy at \( z\sim7 \) on the BPT diagram, using the nebular parameters estimated from the FIR lines.
The upper limits of B14-65666 are predicted to be located in the distribution of star-forming galaxies at \( z\sim2 \).
The location of B14-65666 may be below the \( z\sim2 \) average relation given the weak \NO\ upper limits.
In the case of \(\CPDR = 0.1\), the upper limits are located around the distribution of local galaxies.
Further ALMA statistical observations and rest-frame optical-line observations with JWST will provide opportunities for addressing the high-redshift BPT diagram.

\acknowledgments
We acknowledge Yuichi Harikane for providing us with the Cloudy data.
We thank Ken Mawatari and Takashi Kojima for useful discussions about the details of SED fitting and the \NO\ abundance ratio, respectively.
We wish to thank the referee for constructive and valuable suggestions for improvement.
This research is supported by NAOJ ALMA Scientific Research Grant number 2020-16B and by JSPS KAKENHI Grant Number 17H01114.
TH was supported by Leading Initiative for Excellent Young Researchers, MEXT, Japan (HJH02007).
EZ acknowledges funding from the Swedish National Space Agency.
This paper makes use of the following ALMA data: ADS/JAO.ALMA\#2019.1.01491.S.
ALMA is a partnership of ESO (representing its member states), NSF (USA), and NINS (Japan), together with NRC (Canada), MOST and ASIAA (Taiwan), and KASI (Republic of Korea), in cooperation with the Republic of Chile.
The Joint ALMA Observatory is operated by ESO, AUI/NRAO, and NAOJ.
This research has made use of NASA’s Astrophysics Data System.

\software{CASA \citep{McMullin.J:2007a}, NumPy \citep{Harris.C:2020a},
  SciPy \citep{Virtanen.P:2020a}, IPython \citep{Perez.F:2007a},
  Matplotlib \citep{Hunter.J:2007a}, Astropy \citep{Astropy-Collaboration:2013a, Astropy-Collaboration:2018a},
  APLpy \citep{Robitaille.T:2012a}, \textsc{Cloudy} \citep{Ferland.G:2017a}, PyNeb \citep{Luridiana.V:2015a}}

\bibliographystyle{aasjournal}

\appendix

\section{Results of the radiative equilibrium fitting depending on \( \beta \)}
\begin{figure}[t]
    \epsscale{0.4}
    \plotone{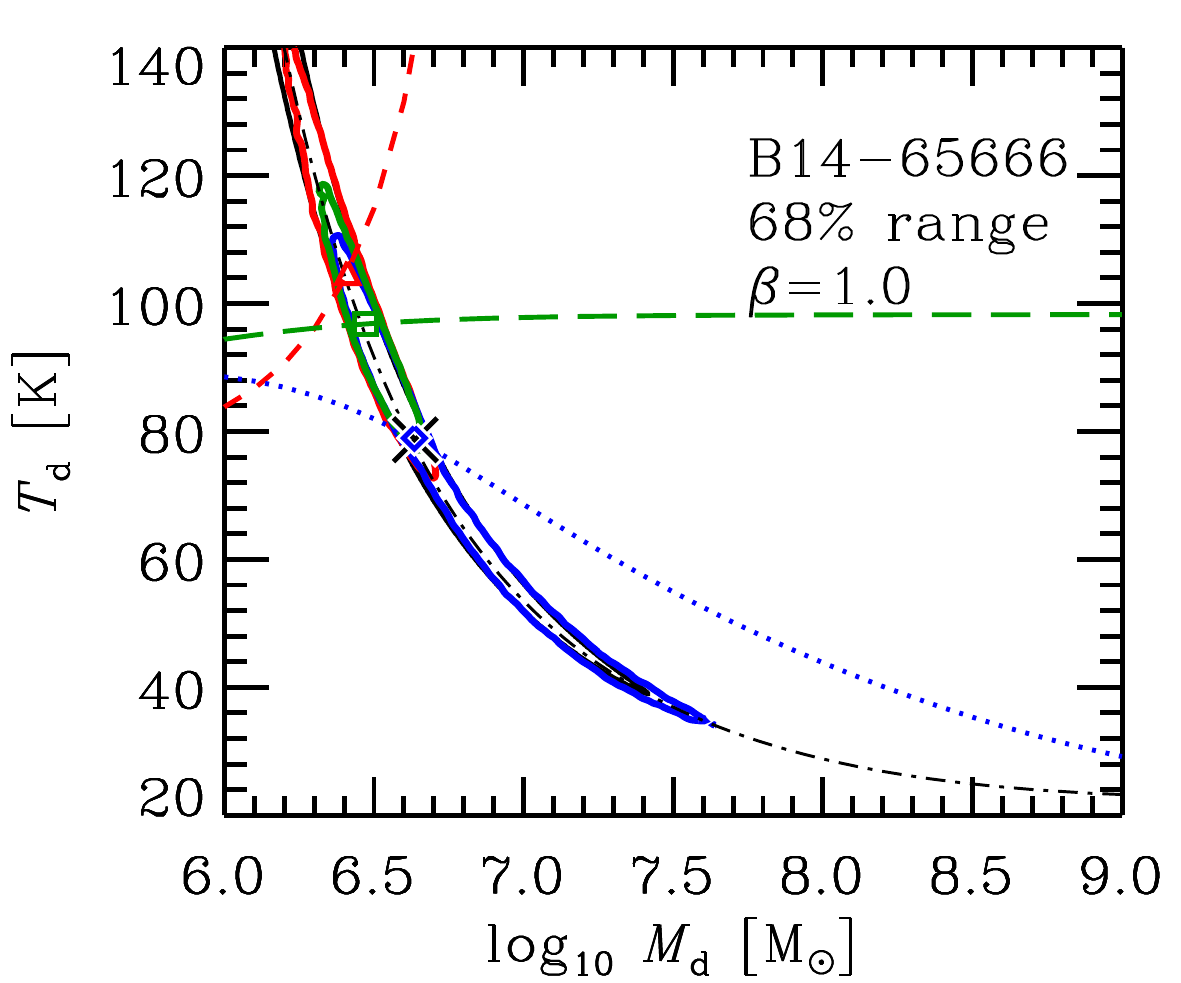}
    \plotone{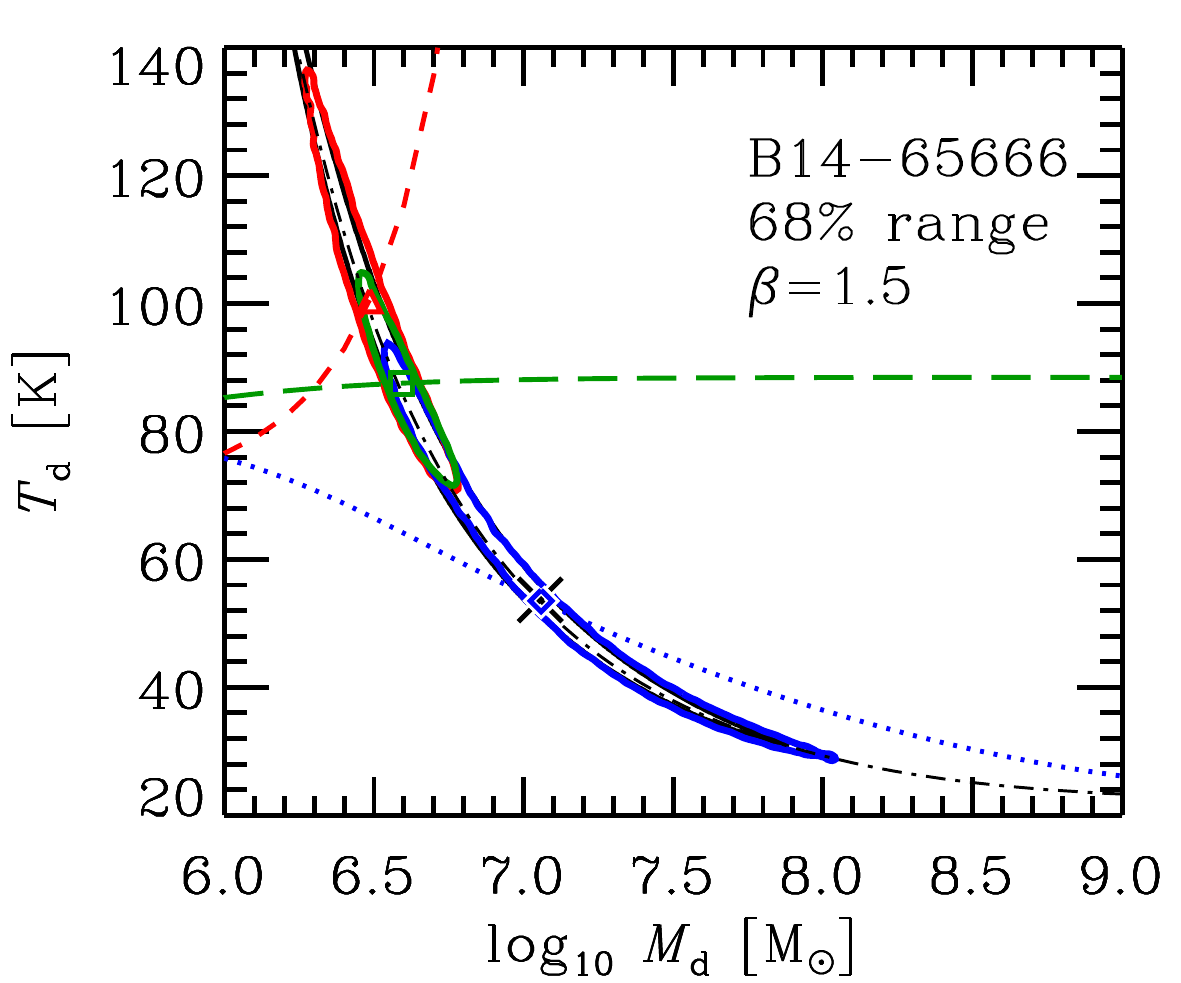}
    \caption{
    Same as Figure~\ref{fig:radeq_MdTd} but for the emissivity indices $\beta=1.0$ and $1.5$.
    }
    \label{fig:radeq_MdTd_ax}
\end{figure}

\begin{figure}[t]
    \epsscale{0.4}
    \plotone{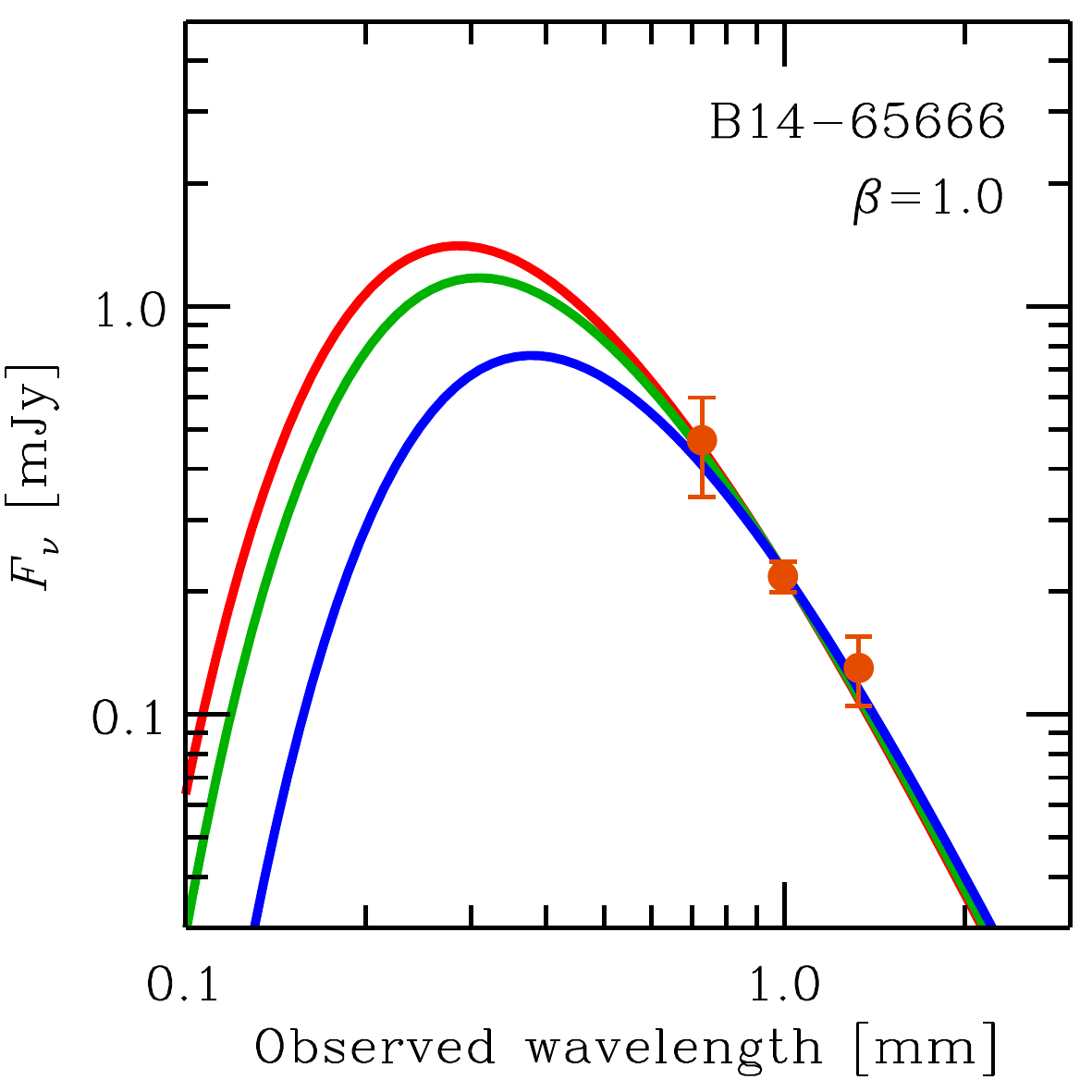}
    \plotone{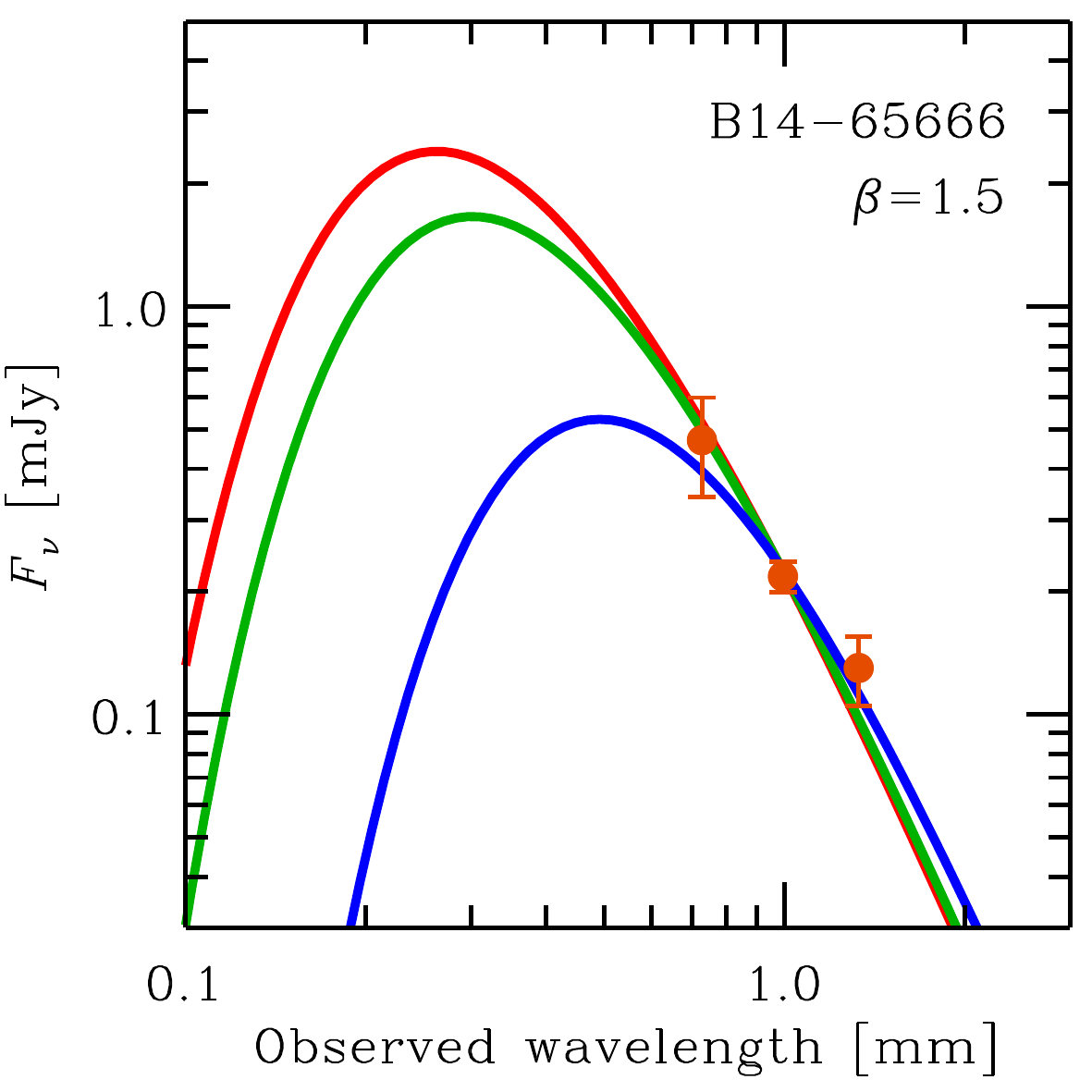}
    \caption{
    Same as Figure~\ref{fig:radeq_MdTd} but for the emissivity indices $\beta=1.0$ and $1.5$.
    }
    \label{fig:radeq_SED_ax}
\end{figure}

\begin{figure}[t]
    \epsscale{0.4}
    \plotone{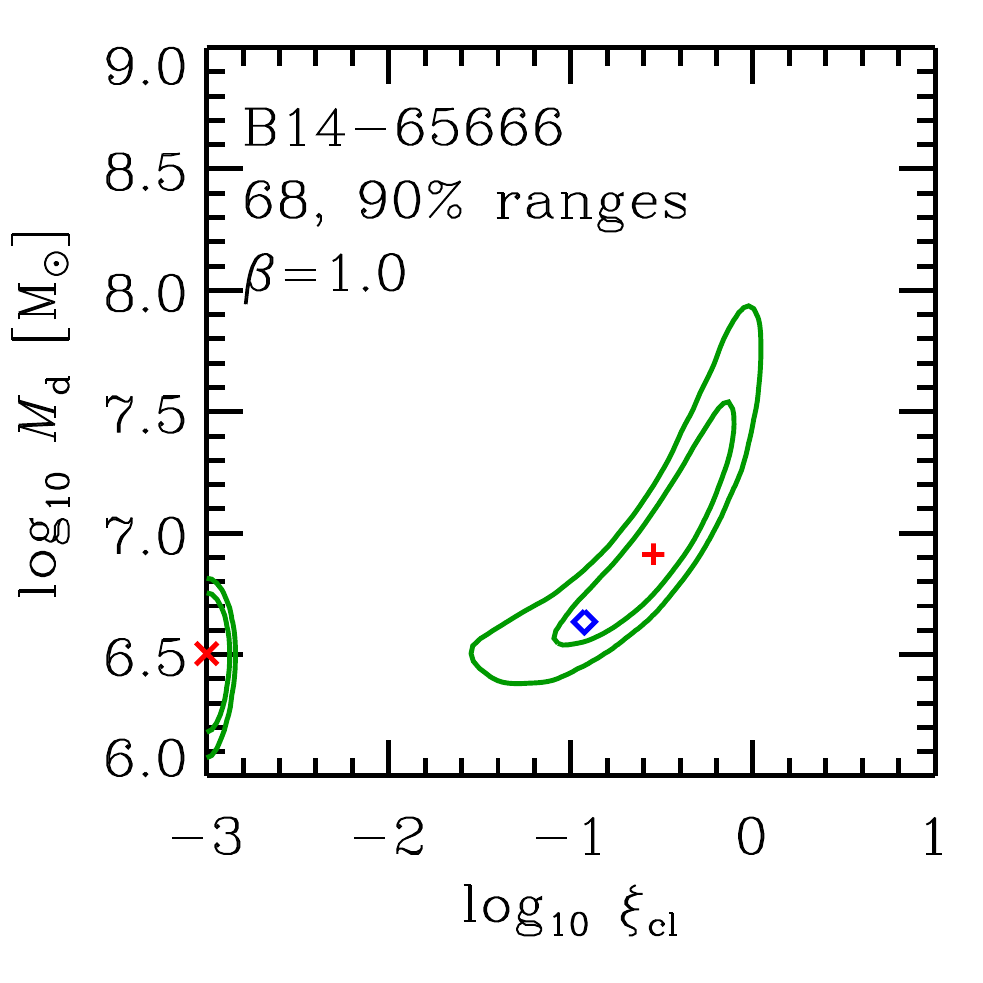}
    \plotone{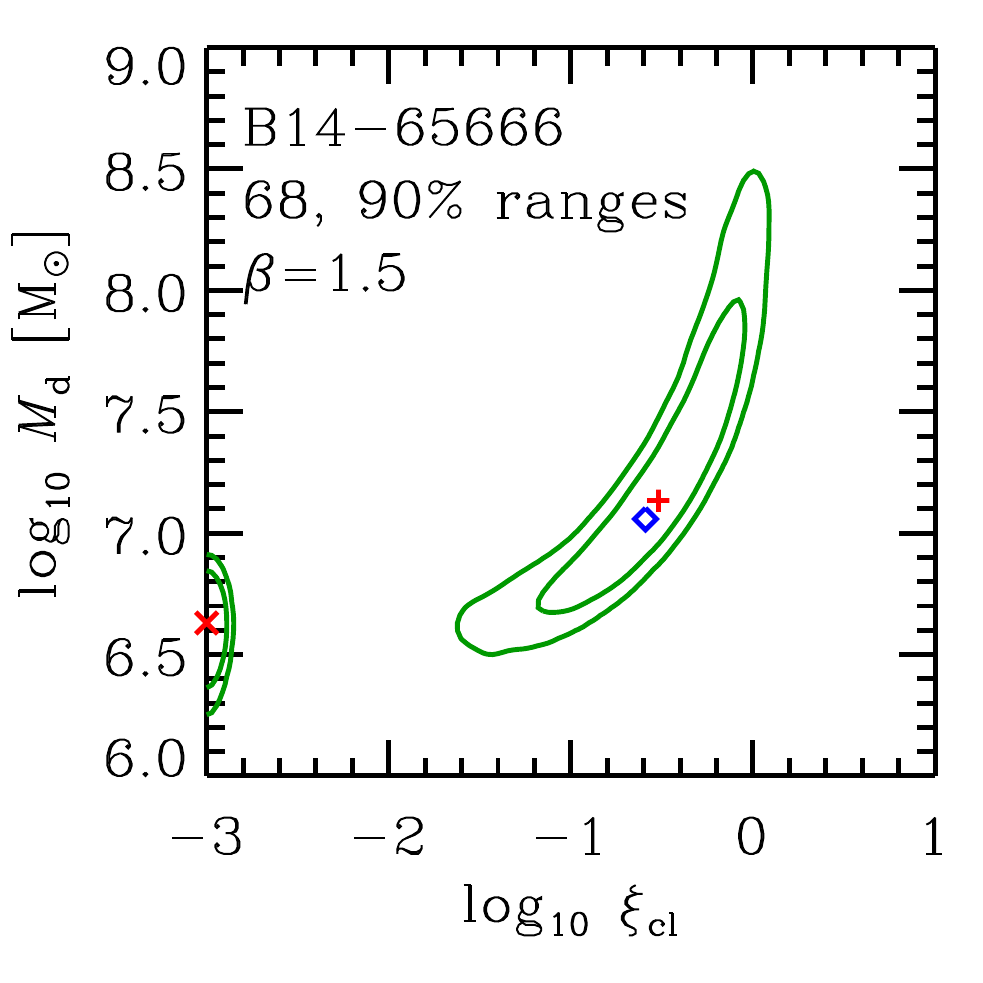}
    \caption{
    Same as Figure~\ref{fig:radeq_MdTd} but for the emissivity indices $\beta=1.0$ and $1.5$. In these cases, the crosses are the highest peaks, and the plus signs are the second highest peaks.
    }
    \label{fig:radeq_Mdclp_ax}
\end{figure}

FIR SED fitting depends on the assumed emissivity index \( \beta \).
The results for \( \beta=2.0 \) is shown in the main text.
Figure~\ref{fig:radeq_MdTd_ax} shows the fitting results of the radiative equilibrium algorithm and Figure~\ref{fig:radeq_SED_ax} shows the best-fit FIR SEDs in the case of \(\beta=1.0\) and \( 1.5 \).
Figure~\ref{fig:radeq_Mdclp_ax} shows the distribution of the solutions in the clumpy geometry for \( \beta=1.0 \) and \( 1.5 \).
Table~\ref{tab:MdTdsummary} summarizes the results of FIR SED fitting, including the modified blackbody fitting.

\begin{deluxetable*}{lccccccc}
    \tablenum{A1}
    \tablecaption{A summary of the FIR SED fitting results.\label{tab:MdTdsummary}}
    \tablewidth{0pt}
    \tablehead{
    \colhead{} & \colhead{} & \colhead{} & \colhead{$T_{\rm d}$} & \colhead{} & \colhead{} & \colhead{} & \colhead{$SFR_{\rm IR}$} \\
    \colhead{Cases} & \colhead{$\chi^2$} & \colhead{d.o.f.} & \colhead{(K)} & \colhead{$\log_{10}(M_{\rm d}/{\rm M}_\odot)$} & \colhead{$\log_{10}\xi_{\rm cl}$} & \colhead{$\log_{10}(L_{\rm IR}/{\rm L}_\odot$)} & \colhead{(M$_\odot$ yr$^{-1}$)} \\
    }
    \decimalcolnumbers
    \startdata
    \multicolumn{6}{l}{Modified blackbody fitting} \\
    $\beta=2.0$ & 1.45 & 1 & 40.8 (29.2--61.5) & 7.45 (6.96--8.06) & --- & 11.62 (11.35--12.19) & 72 \\
    $\beta=1.5$ & 1.07 & 1 & 53.7 (36.5--95.5) & 7.06 (6.52--7.57) & --- & 11.80 (11.39--12.64) & 110 \\
    $\beta=1.0$ & 0.74 & 1 & 78.7 (47.4--255) & 6.64 (5.86--7.15) & --- & 12.09 (11.50--13.86) & 210 \\
    \hline
    \multicolumn{6}{l}{Radiative equilibrium fitting, shell model} \\
    $\beta=2.0$ & 4.53 & 2 & 96.6 (79.8--131) & 6.55 (6.33--6.71) & --- & 12.97 (12.62--13.55) & 1600 \\
    $\beta=1.5$ & 2.19 & 2 & 100 (82.6--137) & 6.49 (6.27--6.64) & --- & 12.72 (12.41--13.25) & 900 \\
    $\beta=1.0$ & 0.88 & 2 & 105 (86.0--142) & 6.41 (6.20--6.56) & --- & 12.49 (12.21--12.95) & 530 \\
    \hline
    \multicolumn{6}{l}{Radiative equilibrium fitting, homogeneous model} \\
    $\beta=2.0$ & 3.62 & 2 & 79.7 (71.8--91.5) & 6.71 (6.59--6.81) & --- & 12.63 (12.44--12.87) & 730 \\
    $\beta=1.5$ & 1.85 & 2 & 87.5 (78.0--102) & 6.59 (6.47--6.69) & --- & 12.50 (12.32--12.75) & 540 \\
    $\beta=1.0$ & 0.82 & 2 & 96.8 (85.2--115) & 6.47 (6.34--6.58) & --- & 12.37 (12.19--12.62) & 400 \\
    \hline
    \multicolumn{6}{l}{Radiative equilibrium fitting, clumpy model} \\
    $\beta=2.0$ & 1.45 & 1 & 40.7 (27.7--64.4) & 7.45 (6.91--8.21) & $-0.43$ ($-1.09$--$-0.11$) & 11.62 (11.42--12.45) & 72 \\
    $\beta=1.5$ & 1.07 & 1 & 53.5 (35.0--82.5) & 7.06 (6.64--7.65) & $-0.59$ ($<-0.19$) & 11.79 (11.40--12.43) & 110 \\
    $\beta=1.0$ & 0.74 & 1 & 79.0 (45.8--99.8) & 6.63 (6.44--7.20) & $-0.93$ ($<-0.31$) & 12.09 (11.48--12.42) & 210 \\
    \enddata
    \tablecomments{
    The values in the parentheses indicate the 68 percent ranges around the best-fit values. For the modified blackbody fitting, both dust temperature, $T_{\rm d}$, and dust mass, $M_{\rm d}$, are the fitting parameters. For the radiative equilibrium fitting, $T_{\rm d}$ and $M_{\rm d}$ are connected and only one of them is the actual fitting parameter. In the clumpy model, another fitting parameter --- the clumpiness, $\xi_{\rm cl}$ --- is introduced. The IR luminosity, $L_{\rm IR}$, is a derived quantity obtained from the integral of the modified blackbody function over the entire frequency range. The IR star formation rate, $SFR_{\rm IR}$, is derived from $L_{\rm IR}$ with the conversion of Kennicutt (1998).
    }
\end{deluxetable*}

\end{document}